\documentclass[[UTF8]{revtex4-1}
\usepackage[utf8]{inputenc}
\usepackage{graphicx}
\usepackage{dcolumn}
\usepackage{bm}
\usepackage{amsmath}
\usepackage{amssymb}
\usepackage{latexsym}
\usepackage{epsfig}
\usepackage{amsbsy}
\usepackage{array}
\usepackage{setspace}
\usepackage{mathrsfs}
\usepackage{geometry}
\usepackage{caption}

\begin{document}

\thispagestyle{empty}
	
\title{The high-order nonrelativistic Hamiltonian of electromagnetic system}
\author{Wanping Zhou}
\affiliation{School of Physics and Telecommunications, Huanggang Normal University, Huanggang, China, 438000}
\author{Xuesong Mei} 
\affiliation{Innovation Academy for Precision Measurement Science and Technology,
Chinese Academy of Sciences, Wuhan, China, 430072}
\author{Haoxue Qiao \footnote{Email: qhx@whu.edu.cn}}
\affiliation{School of Physics and Technology, Wuhan University, Wuhan, 430000 China}

\begin{abstract}
The nonrelativistic Hamiltonians of scalar, spinor and vector particles in the electromagnetic field are studied by applying the Douglas-Kroll-Hess approach. Their relativistic Hamiltonians are expanded on the potential, and the Hamiltonians containing one- and two-photon potentials are derived. The nonrelativistic Hamiltonians up to $m\alpha^8$ order are obtained by applying Taylor expansion on momentum, and the result of spin-1/2 spinor is coincided with the result obtained by applying scattering matching approach in the Ref.~[Phys. Rev. A {\bf 100}, 012513 (2019)]. Then, the singularities in Hamiltonian of Coulomb systems are separated out and cancelled. The regularized Hamiltonian up to $m\alpha^8$ order for scaler and electron in Coulomb field are obtained. The numerical results of relativistic corrections are coincided with the relativistic theory. The regularized Hamiltonian up to $m\alpha^6$ for multi-electrons in Coulomb field are also derived.
\end{abstract}


\maketitle

\newpage

 \section{INTRODUCTION}

The nonrelativistic few-body atomic and molecular systems like hydrogen, helium, lithium and hydrogen molecular ions, are widely studied. Because of the astonishing accuracy of spectroscopy measurements, those systems are advised to determine the fundamental physical constants and search the hint of the new physics beyond the standard model \cite{PhysRevLett.113.023004,PhysRevLett.118.063001,RevModPhys.88.035009,RevModPhys.90.025008}. In the aspects of theories, the nonrelativistic energy of systems can be calculated with very high accuracy by applying Hyllerass algorithm \cite{ Hylleraas1928,Hylleraas1929,YAN199696,PhysRevA.65.054501,PhysRevA.85.052513}. The relativistic and radiative corrections are calculated by applying perturbation theory. The key point is deriving the formula of the relativistic and radiative corrections.

There are two famous methods obtaining the relativistic and radiative corrections to the nonrelativistic systems. One is an effective field theory proposed by Pachucki \emph{et.al}. Its low-energy effective Hamiltonian is the Foldy-Wouthuysen (FW) transformed Hamiltonian \cite{PhysRev.78.29,PhysRevA.82.052520}. The effective currents of electron coupling with electromagnetic field can be extracted from the FW Hamiltonian, and the high-order interactions between the electrons and the self-energy correction can be derived by using the currents\cite{PhysRevLett.104.070403,PhysRevA.71.012503,PhysRevA.94.052508,PhysRevA.95.012508,PhysRevA.72.062102,PhysRevA.74.022512}. Another method is the nonrelativistic quantum electrodynamics~(NRQED), proposed by W.~Caswell and G.~Lepage \cite{CASWELL1986437} and successfully be applied on positronium and muonium\cite{PhysRevD.53.4909,PhysRevD.55.7267}. The low-energy effective Hamiltonian of NRQED consists of all possible local interactions satisfying the required symmetries of quantum electrodynamics~(QED). And the coefficients of the interaction terms of the effective Hamiltonian are fixed by scattering matching approach\cite{CASWELL1986437}: the scattering amplitude of the NRQED must be coincided with the corresponding scattering amplitude of the QED at given energy scale. The great difference between the two method is that their effective Hamiltonians are different.

Both the high-order energy corrections and operator of Hamiltonians can be expanded with the fine-structure constant $\alpha$ \cite{IEIDES200163,PhysRevD.53.4909,PhysRevD.55.7267}.  For helium, the energy correction for helium up to $m\alpha^6$ and the fine-structure correction up to $m\alpha^7$ have been obtained \cite{PhysRevA.71.012503,PhysRevLett.104.070403,PhysRevA.100.042510}. The energy correction up to order of $m\alpha^7$ allowed one to directly extract the nuclear charge radii \cite{PhysRevA.95.062510, PhysRevA.103.042809}. The values of nuclear charge radii extracted from transitions $2^3S-2^3P$ and $2^3S-2^1S$ have notable discrepancy~\cite{vanrooij2011frequency, rengelink2018precision, PhysRevLett.119.263002}. For hydrogen molecular ions, ionization energies and fundamental vibrational transitions are calculated up to $m\alpha^7$ order \cite{PhysRevLett.118.233001}. The $m\alpha^8$ loop corrections are estimated, and the relativistic corrections haven't derived. The $m\alpha^8$ corrections could be a very big challenge for theoretical study in nowadays. Its result will improve both theoretical accuracy of spectroscopy experiment and fundamental physical constants \cite{PhysRevLett.104.070403,PhysRevA.95.062510,RevModPhys.80.633}. One purpose of this work is deriving the $m\alpha^8$ order FW hamiltonian. Another purpose is testing the equivalence of Pachucki's effective field theory and NRQED by comparing the FW Hamiltonian and NRQED Hamiltonian. The FW Hamiltonian of spinor up to $m\alpha^8$ was obtained by Ref.~\cite{MEIXueSong63102}. In Ref. \cite{PhysRevA.100.012513}, the $m\alpha^8$ order NRQED Hamiltonian was obtained by using scattering matching approach at the tree diagram level. It is coincided with the FW Hamiltonian at $m\alpha^6$ order. At $m\alpha^8$ order we find that there may be some imperceptible mistakes in Ref.~\cite{MEIXueSong63102}. It should be rechecked independently.

The FW transformation is unitary transformation that can decouple the positive energy parts and the negative energy part of particles in the low-energy region. By applying the FW transformation, the relativistic Hamiltonians are expanded on the momentum and potential of charged particles. The traditional approach of FW transformation becomes rather tedious in the higher-order. The Douglas-Kroll-Hess (DKH) approach  \cite{DOUGLAS197489,PhysRevA.33.3742,LUBER2009205,doi:10.1063/1.1768160,doi:10.1063/1.1818681} is the alternative approach of FW transformation. The relativistic Hamiltonians are expanded on the potential of charged particles. It is two-component relativistic Hamiltonian for fermion. However, the DKH Hamiltonian is too complicated to be combined with Hyllerass algorithm \cite{ Hylleraas1928,Hylleraas1929,YAN199696,PhysRevA.65.054501,PhysRevA.85.052513}, which is the most accurate method for calculating the electrons' correlation. The DKH Hamiltonian should be expanded on momentum to obtain the FW Hamiltonian. This DHK-FW approach is easier than the FW approach. In this work, DHK-FW method is applied to derive the $m\alpha^8$ order FW Hamiltonian of spin-1/2 case and compare them with the results obtained by scattering approaching \cite{PhysRevA.100.012513,CASWELL1986437}. Furthermore, the effective Hamiltonians for spin-0 and spin 1 cases also derived. It is essential for deriving the interactions between electrons and integer spin heavier particle, such as nucleus, muon or vector meson. The higher-order interactions between the electrons and integer spin heavier particle can be derived by using the currents extracted from the FW Hamiltonian. It is essential for deriving the recoil effect of spin heavier particle and isotopic shift.

The FW Hamiltonian is singular at high order. For example, the relativistic kinetic energy $\langle p^{6}\rangle$ is proportional to  $\langle \dfrac{\delta^3(r)}{r}\rangle\rightarrow\infty$ for s state of hydrogen, and the expectation value of Hamiltonian is always divergent. The second perturbation term such as $\langle p^{4}\frac{Q}{E_{0}-H_{0}}p^{4}\rangle$ also contains divergent part. All the divergent parts in the higher-order energy corrections should be cancelled exactly. Because the FW Hamiltonian is unitary equivalent to the relativistic Hamiltonian, their eigenvalue are equal and finite. It is essential to separate and cancel the singular terms from original nonrelativistic Hamiltonian, and then deriving the calculable formula. In this work, we study the Coulomb Hamiltonian and cancel the divergence by applying transformation. Then, we obtain an equivalent formula of energy correction. In the transformed formula, Most singularities of the FW Hamiltonian are cancelled directly. The regularized Coulomb Hamiltonians up to $m\alpha^8$ order for scaler and electron are obtained. The regularized Hamiltonian up to $m\alpha^6$ for multi-electrons in Coulomb field are also derived. The singularities always appear in the high-order correction, such as the relativistic correction, radiative correction and photon-exchanging interactions. It is essential to cancel all the divergences. The singularities of the total Hamiltonian up to $m\alpha^6$ order were cancelled together\cite{PhysRevA.74.022512}. Our study will indicate the singularities can be cancelled part by part. It also simplifies the verifying the correctness of the theoretical results in the higher-order. 

In Sec.~\ref{NRHamiltonian}, we derive the FW Hamiltonians of the spin-1/2, spin-0 and spin-1 particles separately by using DKH-FW approach. In Sec.~\ref{ElimS}, The singularities of Coulomb Hamiltonian are studied. We derive the $m\alpha^8$ order  regularized Coulomb Hamiltonians of the scaler and electron, and $m\alpha^6$ order regularized Coulomb Hamiltonians of multi-electron systems. In Sec.~\ref{checking}, the  derived formulas are checked. The relativistic corrections to energy of scalar and electron in Coulomb field are calculated by using the regularized Coulomb Hamiltonians. The results are coincided with energy of relativistic theory. Conclusion is given in Sec.~\ref{conclusion}. 

\section{The nonrelativistic Hamiltonians of scalar, spinor and vector particles}\label{NRHamiltonian}

\subsection{Spin-1/2 case} \label{spin1/2}

\subsubsection{Spinor in electromagnetic field}
The relativistic spin-1/2 fermion in the electromagnetic field is described by the Dirac Hamiltonian
$H_{F}=\beta m+\vec{\alpha} \cdot \vec{\pi}+A^{0}$,
where
$\alpha^{i}=\left(\begin{array}{cc}\bm{0}_{2 \times 2} & \bm{\sigma}^{i} \\
\bm{\sigma}^{i} & \bm{0}_{2 \times 2}\end{array}\right),
\beta=\left(\begin{array}{cc}\bm{1}_{2 \times 2} & \bm{0}_{2 \times 2} \\
 \bm{0}_{2 \times 2} & -\bm{1}_{2 \times 2}\end{array}\right)$,
 $\sigma^{i}$ is Pauli matrix, and
 $\pi_{\mu}=p_{\mu}-A_{\mu}$. In this work, the velocity of light and the mass and charges of particles are chosen as unit one, and the form factor and multi-pole terms are neglected for simplification.

In the rest frame, the two upper components of spinor wave function are the freedom of positive energy particle, and the lower components are the freedom of negative energy particle. The positive part couples with the negative part due by the off-diagonal matrices like $\vec{\alpha} \cdot \vec{\pi}$. These off-diagonal matrices are defined as odd terms, and the diagonal parts of the Hamiltonian are defined as even terms. By applying Foldy-Wouthuysen transformation, the odd terms become negligible higher-order terms, and the positive part can be decoupled with the negative parts in the low-energy region.

The FW Hamiltonian $\hat H_{F W}=e^{i S}\left(\hat H-i \partial_{t}\right) e^{-i S}$ can be derived by commutator form \cite{PhysRevA.82.052520}
\begin{equation}\label{FWperturbation}\begin{aligned}
H_{F W} &=\sum_{n=0} \frac{1}{n !}\left[(i S)^{(n)},\left(H-i \partial_{t}\right)\right]
+i \partial_{t},
\end{aligned}\end{equation}
where recurrence relations $[{A}^{(n+1)},{B}] =[{A},[{A}^{(n)},{B}]]$ and $[{A}^{(0)},B] =B$ are used. According to the commutator, the odd terms can be suppressed to smaller contributions and make the positive and negative part no longer coupling with each other up to the accuracy of requirements. However, with the increase of the order of magnitude, this process will become bloated and complicated. Because the decoupling is synchronized with the momentum expansion and potential expansion in the traditional FW transformation. 

By DKH approach, the Hamiltonian is expanded with the electromagnetic potential. The first transformation will produce the exact form of the relativistic energy $\sqrt{p^2+m^2}$ of free particle, and the reminder odd terms are eliminated by applying subsequent transformation Eq.(\ref{FWperturbation}). The Hamiltonian is expanded in the power of the potential.

For spin-1/2 fermion, the first transformation is $e^{i S_{F}}=c_{F}+i \beta s_{F}$, where
$c_{F}=\frac{\sqrt{m+E_{T}}}{\sqrt{2 E_{T}}},
s_{F}=\frac{-i(\alpha \cdot \pi)}{\sqrt{2 E_{T}\left(E_{T}+m\right)}},
S_{F}=-i \frac{1}{2} \beta \tan^{-1} \frac{\alpha \cdot \pi}{m}$
and
$E_{T}=\sqrt{m^{2}+(\alpha \cdot \pi)^{2}}$.
The transformed Hamiltonian can be written as
\begin{equation}\label{FWFermion}
H_{F}^{\prime}=e^{i S_{F}}\left(H_{F}-i \partial_{t}\right) e^{-i S_{F}}=\beta E_{T}+H_{1 \gamma}^{\prime}+O^{\prime}.
\end{equation}
The first two terms $\beta E_T, H'_{1\gamma}$ are even terms, $O'$ represents the odd terms. The relativistic kinetic energy $\beta E_{T}$ is exact here. The even one-photon terms $H_{1 \gamma}^{\prime}$ only contain one scalar potential term 
\begin{equation}
H_{1 \gamma}^{\prime}=c_{F}\left(A^{0}-i \partial_{t}\right) c_{F}-s_{F}\left(A^{0}-i \partial_{t}\right) s_{F},
\end{equation}
and the one-photon odd terms $O^{\prime}$ are
\begin{equation}
O^{\prime}=i\beta s_{F}\left(A^{0}-i \partial_{t}\right) c_{F}-i\beta c_{F}\left(A^{0}-i \partial_{t}\right) s_{F}.
\end{equation}
 
The odd one-photon terms can be eliminated by applying another unitary transformation $e^{i S_{F}^{\prime}}$. The first-order Eq.(\ref{FWperturbation}) is $\left[i S_{F}^{\prime}, \beta E_{T} \right]-\partial_{t} S_{F}^{\prime}+O^{\prime}=0$, and $S_{F}^{\prime}$ can be derived by the recurrence relation,
\begin{equation}
S_{F}^{\prime}=-\frac{1}{2 m} i \beta \sum_{n=0}\left\{\left(\frac{i \beta \partial_{t}-(\sqrt{\pi^{2}+m^{2}}-m)}{2 m}\right)^{(n)}, O^{\prime}\right\},
\end{equation}
the curly braces is anti-commutator.

The second transformed Hamiltonian is
\begin{equation}\label{Fermion_Hamiltonian2}
H^{\prime \prime}_{F}=\beta E_{T}+H_{1 \gamma}^{\prime}+H_{2 \gamma}^{\prime}+\ldots,
\end{equation}
where
\begin{equation}\label{H2gamma}
H_{2 \gamma}^{\prime}=\frac{1}{2}\left[i S^{\prime}_{F}, O^{\prime}\right]=\frac{1}{2 m} \beta \sum_{n=0}\left\{\left\{\left(\frac{i \beta \partial_{t}-(\sqrt{\pi^{2}+m^{2}}-m)}{2 m}\right)^{(n)}, O^{\prime}\right\}, O^{\prime}\right\},
\end{equation}
are even two-photons terms, which contains two scalar potential. The odd two-photons terms and the terms containing more than two scalar potential are neglected here.

The Hamiltonian Eq.(\ref{Fermion_Hamiltonian2}) can be expanded on momentum to arbitrary order. In the Coulomb bound system, the order of magnitude of electromagnetic potential is counted as $\langle p^{i} \rangle \sim \alpha$,  $\langle A^{0} \rangle \sim \alpha^2$ and $\langle A^{i} \rangle \sim \alpha^3$. Then Eq.(\ref{Fermion_Hamiltonian2}) is expanded on momentum or the fine-structure constant $\alpha$. The effective Hamiltonian (FW) up to $m\alpha^8$ is
\begin{equation}\begin{aligned}
H^{\prime \prime}_{F}=&\beta E_{T}+A^0
-\frac{i}{8}[\widetilde{\pi}, \widetilde{E}]
-\frac{i}{128}\left[\widetilde{\pi}^{2},\{\widetilde{\pi}, \widetilde{E}\}\right]
+\frac{3 i}{32}\left[\widetilde{\pi}^{3}, \widetilde{E}\right]
+\frac{1}{16} \beta\{\widetilde{E}, \widetilde{E}\}\\
&+\frac{11 i}{1024}\left[\widetilde{\pi}^{4},\{\widetilde{\pi}, \widetilde{E}\}\right]
-\frac{31 i}{512}\left[\widetilde{\pi}^{5}, \widetilde{E}\right]
-\frac{9 i}{512}\left[\widetilde{\pi}^{3}, \widetilde{\pi}^{2} \widetilde{E}+\widetilde{E} \widetilde{\pi}^{2}
+\widetilde{\pi} \widetilde{E} \widetilde{\pi}\right]\\
&+\frac{i}{32}\left[\partial_{t} \widetilde{E}, \widetilde{E}\right]
-\frac{1}{32} \beta\left\{\widetilde{E},\left(2 \widetilde{\pi}^{2} \widetilde{E}
+2 \widetilde{E} \widetilde{\pi}^{2}
+\widetilde{\pi} \widetilde{E} \widetilde{\pi}\right)\right\}
+o\left(m \alpha^{8}\right),
\end{aligned}\end{equation}
where the tilde is abbreviation defined as $\widetilde{F}=\vec{\sigma} \cdot \vec{F} $. This result is coincided with the result obtained by scattering matching approach in Ref.~\cite{PhysRevA.100.012513}.

It is beneficial to compare these two methods. In Ref.~\cite{PhysRevA.100.012513}, one-photon terms are derived by expanding fermion-photon scattering amplitude $\bar{u}(p') \gamma^{\mu} u(p) A_{\mu}$ on 3-momentums, where $u(p)$ is the free positive-energy Dirac spinor. The expansion of $\bar{u}(p') \gamma^{\mu} u(p) A_{\mu}$ is $\psi^{\dagger} H^{\prime }_{1\gamma} \psi$, and $H^{\prime }_{1\gamma}$ are coincided with one-photon potential even terms given above. There are another two fermion-photon scattering processes $\bar{v}(p') \gamma^{\mu} u(p) A_{\mu}$ and $\bar{u}(p') \gamma^{\mu} v(p) A_{\mu}$. They are the coupling terms between photon $ A_{\mu}$, positive-energy states $u(p')$ and negative-energy states $v(p)$. It is corresponding to the one-photon odd terms $O^{\prime}$. In the low-energy region, free negative state is decoupled. But the contribution of negative-energy parts to the intermediate state should be considered. The positive-energy state can transit to negative-energy intermediate state, and transits back to positive-energy state. Because the energy of intermediate state $E\simeq 2 m c^2 \gg E_{bound}$, the Green function of the intermediate state can be replaced with $\frac{1}{2m}$. The contribution of negative-energy intermediate state is proportional to the product of $\bar{v}(p') \gamma^{\mu} u(p) A_{\mu}$, $\bar{u}(p') \gamma^{\mu} v(p) A_{\mu}$ and  $\frac{1}{2mc^2}$. It is corresponding to the product of $O'^2$ and $\frac{1}{2m}$ in the Eq.(\ref{H2gamma}). Through comparison of above results and method, it can be concluded that: the DKH-FW approach is equivalent to scattering matching at $m\alpha^8$ order.

\subsubsection{The nonrelativistic approximation of Dirac-Coulomb Hamiltonian}

The relativistic energy can be calculated by using the Dirac-Coulomb(DC) Hamiltonian.
The DC Hamiltonian is 
\begin{equation} \label{DCE}
\begin{aligned}
H_{DC}=\sum_{a} \left( \alpha_a\cdot p_a+\beta_a m_a\right)+V_{C},
\end{aligned}
\end{equation}
where the Coulomb potential is
\begin{equation} 
\begin{aligned}
V_{C}=-\sum_{i,a}\frac{Z_{i}}{r_{ia}}+\sum_{a<b}\frac{1}{r_{ab}}.
\end{aligned}
\end{equation}
It contains both the electron-electron (subscript i,a) and electron-nucleus interaction (subscript a,b).

The nonrelativistic approximation of Dirac-Coulomb Hamiltonian can be obtained by using the similar method. In this subsection, we will study the two electrons systems, and the results can be extended to the general case.

The first transformation also produces the exact form of the relativistic energy $\sqrt{p^2+m^2}$. It is  the same as the form in Eq.(\ref{FWFermion})
\begin{equation} 
H'_{DC}=e^{iS_{2}}e^{iS_{1}}H_{DC}e^{-iS_{2}}e^{-iS_{1}}
=\beta_{1}\sqrt{m^{2}_{1}+p^{2}_{1}}+\beta_{2}\sqrt{m^{2}_{2}+p^{2}_{2}}
+V'_{ee}+V'_{eo}+V'_{oe}+V'_{oo},
\end{equation}
where
$e^{i S_{a}}=c_{a}+i \beta s_{a}$,
$c_{a}=\frac{\sqrt{m_{a}+E_{a}}}{\sqrt{2 E_{a}}},
s_{a}=\frac{-i(\alpha_{a} \cdot p_{a})}{\sqrt{2 E_{a}\left(E_{a}+m_{a}\right)}},
S_{a}=-i \frac{1}{2} \beta_{a} \tan^{-1} \frac{\alpha_{a} \cdot p_{a}}{m_{a}}$,
$E_{a}=\sqrt{m^{2}_{a}+p_{a}^{2}}$ 
and 
\begin{equation} \label{DCE}
\begin{aligned}
V'_{ee}&=c_{1}c_{2}V_{C}c_{1}c_{2}
-c_{1}i\beta_{2}s_{2}V_{C}c_{1}i\beta_{2}s_{2}
-i\beta_{1}s_{1}c_{2}V_{C}i\beta_{1}s_{1}c_{2}
+i\beta_{1}s_{1}i\beta_{2}s_{2}V_{C}i\beta_{1}s_{1}i\beta_{2}s_{2},
\\
V'_{eo}&=-c_{1}c_{2}V_{C}c_{1}i\beta_{2}s_{2}
+c_{1}i\beta_{2}s_{2}V_{C}c_{1}c_{2}
+i\beta_{1}s_{1}c_{2}V_{C}i\beta_{1}s_{1}i\beta_{2}s_{2}
-i\beta_{1}s_{1}i\beta_{2}s_{2}V_{C}i\beta_{1}s_{1}c_{2}
\\
V'_{oe}&=-c_{1}c_{2}V_{C}i\beta_{1}s_{1}c_{2}
+i\beta_{1}s_{1}c_{2}V_{C}c_{1}c_{2}
+c_{1}i\beta_{2}s_{2}V_{C}i\beta_{1}s_{1}i\beta_{2}s_{2}
-i\beta_{1}s_{1}i\beta_{2}s_{2}V_{C}c_{1}i\beta_{2}s_{2}
\\
V'_{oo}&=c_{1}c_{2}V_{C}i\beta_{1}s_{1}i\beta_{2}s_{2}
-c_{1}i\beta_{2}s_{2}V_{C}i\beta_{1}s_{1}c_{2}
-i\beta_{1}s_{1}c_{2}V_{C}c_{1}i\beta_{2}s_{2}
+i\beta_{1}s_{1}i\beta_{2}s_{2}V_{C}c_{1}c_{2}
.
\end{aligned}
\end{equation}
The subscript $eo$ is abbreviated as even and odd. $V_{eo}$ is even for first particle and odd for second particle, etc. 

The odd terms or odd-even-mixing terms should be eliminated by applying another transformation Eq.(\ref{FWperturbation})
\begin{equation} 
\begin{aligned}
H''_{DC}&=e^{iS'}H'_{DC}e^{-iS'}
\\&
=\beta_{1}\sqrt{m^{2}_{1}+p^{2}_{1}}+\beta_{2}\sqrt{m^{2}_{2}+p^{2}_{2}}
+V'_{ee}+\frac{1}{2}[iS',V'_{eo}+V'_{oe}+V'_{oo}]+....
\end{aligned}
\end{equation}
The last terms are two-photon terms, and the ellipsis is the multi-photon term, which is $m\alpha^{12}$ order. The $S'$ satisfies the first-order Eq.(\ref{FWperturbation})
\begin{equation} 
[iS',\beta_{1}\sqrt{m^{2}_{1}+p^{2}_{1}}+\beta_{2}\sqrt{m^{2}_{2}+p^{2}_{2}}]
+V'_{eo}+V'_{oe}+V'_{oo}=0.
\end{equation}
It can be solved by using iteration method. The result is
\begin{equation} 
iS'=\frac{\beta_{1}}{2m_{1}}M_{oe}+\frac{\beta_{2}}{2m_{2}}M_{eo}
+\frac{\beta_{1}m_{1}-\beta_{2}m_{2}}{2(m_{1}^2-m_{2}^2)}N_{oo},
\end{equation}
where
\begin{equation} 
M_{oe}=\sum_{n=0}\left(-\frac{\beta_{1}}{2m_{1}}\right)^{n}
[\triangle^{(n)},V'_{oe}],
\end{equation}
\begin{equation} 
M_{eo}=\sum_{n=0}\left(-\frac{\beta_{2}}{2m_{2}}\right)^{n}
[\triangle^{(n)},V'_{eo}],
\end{equation}
\begin{equation} 
N_{oo}=\sum_{n=0}\left(-\frac{\beta_{1}m_{1}-\beta_{2}m_{2}}{2(m_{1}^2-m_{2}^2)}\right)^{n}
[\triangle^{(n)},V'_{oo}],
\end{equation}
and
\begin{equation}
\triangle=\beta_{1}(\sqrt{m^{2}_{1}+p^{2}_{1}}-m_{1})
+\beta_{2}(\sqrt{m^{2}_{2}+p^{2}_{2}}-m_{2}).
\end{equation}

The nonrelativisitc approximation of Dirac-Coulomb Hamiltonian can be obtained by applying momentum expansion. The transformed Hamiltonian up to $m\alpha^6$ order is
\begin{equation} 
\begin{aligned}
H''_{DC}=& \sum_{a}\frac{p^{2}_{a}}{2}+V_{C}
+\sum_{a}\left(-\frac{p^{4}_{a}}{8}
-\frac{1}{8}[\widetilde{p_{a}},[\widetilde{p_{a}},V_{C}]]\right)
\\&
+\sum_{a}\left(\frac{p^{6}_{a}}{16}
+\frac{3}{32}[\widetilde{p_{a}}^3,[\widetilde{p_{a}},V_{C}]]
-\frac{1}{128}[\widetilde{p_{a}}^2,[\widetilde{p_{a}}^2,V_{C}]
-\frac{1}{16}\{[\widetilde{p_{a}},V_{C}],[\widetilde{p_{a}},V_{C}]\}
\right)
\\&
+\sum_{a<b}\left(
\frac{1}{64}[\widetilde{p_{a}},[\widetilde{p_{a}},[\widetilde{p_{b}},[\widetilde{p_{b}},V_{C}]]]]
\right)
+o(m\alpha^6).
\end{aligned}
\end{equation}
The first two terms are Shr$\ddot{o}$dinger Hamiltonian. The third term is the leading order of relativistic corrections, and it is $m\alpha^4$ order. The second and third lines are $m\alpha^6$ order.  Although it is nonrelativisitc Hamiltonian of two electrons system, this result is appropriate to multi-electron atom. It can be proved by using the similar method.

The $m\alpha^8$ order terms of Hamiltonian of two electrons systems are
\begin{equation} 
\begin{aligned}
H''_{DC(8)}=&  
-\frac{5p^{8}_{a}}{128}
-\frac{9}{512}[\widetilde{p_{a}}^3,[\widetilde{p_{a}}^3,V_{C}]]
-\frac{31}{512}[\widetilde{p_{a}}^5,[\widetilde{p_{a}},V_{C}]]
+\frac{11}{1024}[\widetilde{p_{a}}^4,[\widetilde{p_{a}}^2,V_{C}]]
\\&
-\frac{1}{8}\{[\widetilde{p_{a}},V_{C}],
\left(
-\frac{3}{16}[\widetilde{p_{a}}^3,V_{C}]
-\frac{5}{32}\{\widetilde{p_{a}}^2,[\widetilde{p_{a}},V_{C}]\}
+\frac{1}{32}[\widetilde{p_{a}}^2,\{\widetilde{p_{a}},V_{C}\}]
\right)
\}
\\&
+\frac{3}{256}[\widetilde{p_{a}}^3,[\widetilde{p_{a}},[\widetilde{p_{b}},[\widetilde{p_{b}},V_{C}]]]]
+\frac{1}{1024}[\widetilde{p_{a}}^2,[\widetilde{p_{a}}^2,[\widetilde{p_{b}},[\widetilde{p_{b}},V_{C}]]]]
\\&
-\frac{1}{128}\{[\widetilde{p_{a}},V_{C}],
[\widetilde{p_{b}},[\widetilde{p_{b}},[\widetilde{p_{a}},V_{C}]]]
\}
+\frac{1}{128}\{[\widetilde{p_{a}},[\widetilde{p_{b}},V_{C}]],
[\widetilde{p_{a}},[\widetilde{p_{b}},V_{C}]]\}
\\&
-\frac{1}{64}[[\widetilde{p_{a}},V_{C}],
[p_{b}^2,[\widetilde{p_{a}},V_{C}]]],
\end{aligned}
\end{equation}
where the summation $a$ and $a<b$ are omitted.

\subsection{Spin-0 case} \label{spin0}
In the electromagnetic field, the equation of the scalar field is $\left(\pi^{\mu} \pi_{\mu}-m^{2}\right) \phi=0$. The positive energy part and the negative part wave functions are respectively $\theta=\frac{1}{2}\left(\phi+\frac{1}{m} \pi^{0} \phi\right)$ and $\chi=\frac{1}{2}\left(\phi-\frac{1}{m} \pi^{0} \phi\right)$ \cite{PhysRevA.82.052520}, where the $\frac{1}{2}\left(1\pm\frac{1}{m} \pi^{0}\right)$ is the positive$\slash$negative project operator. In the rest reference frame, positive state satisfies $\chi=0$, and negative state satisfies  $\theta=0$. 
 
One can find out that the equation of scalar field equivalent expression is $i \eta \partial_{t} \Phi=H_{s} \Phi$,
where
$\eta=\left(\begin{array}{cc}1 & 0 \\ 0 & -1\end{array}\right),
 \tau=\left(\begin{array}{ll}0 & 1 \\ 1 & 0\end{array}\right)$,
$\Phi=\left(\begin{array}{l}\theta \\ \chi\end{array}\right)$ and the Hamitonian is
\begin{equation}
H_{s}=(1+\tau) \frac{1}{2 m} \vec{\pi}^{2}+m+\eta A_{0}.
\end{equation}
In this Hamiltonian, the positive energy part couple with the negative part through $\tau$. The terms containing $\tau$ is odd. The purpose of FW and DKH transformation is reducing the odd term to be insignificant. The transformations of wave function and Hamiltonian are $\Phi={U} \Phi^{\prime}$ and $H^{\prime}_{FW}=U^{\dagger}\left(H-i \eta \partial_{t}\right) U$. The ${U}=e^{{S}}$ satisfy $U^{\dagger} \eta U=\eta$ , and ${S}$ is a Hermitian operator. Because the normalization condition of the $\Phi$ is $\int \Phi^{\dagger} \eta \Phi d^{3} x=1$.

Let $S=-\frac{\tau}{2} \ln \frac{E_{T}}{m}$, and $E_{T}=\sqrt{\vec{\pi}^{2}+m^{2}}$. The transformed Hamiltonian is
  \begin{equation}
  H_{s}^{\prime}=E_{T}+H_{1 \gamma}^{\prime}+O^{\prime}.
  \end{equation}
The even one-photon terms and odd one-photon terms are
\begin{equation}
H_{1 \gamma}^{\prime}=\eta A_{0}+\frac{1}{2} \eta\left(\left[s_{s},\left[s_{s},\left(A_{0}-i \partial_{t} S\right)\right]\right]-\left[c_{s},\left[c_{s},\left(A_{0}-i \partial_{t} S\right)\right]\right]\right),
\end{equation}
and 
\begin{equation}
O^{\prime}=\frac{1}{2} \eta\left(-\left\{c_{s},\left[s_{s},\left(A_{0}-i \partial_{t} S\right)\right]\right\}+\left\{s_{s},\left[c_{s},\left(A_{0}-i \partial_{t} S\right)\right]\right\}\right),
\end{equation}
where $s_{s}=-\frac{\tau \pi^{2}}{\sqrt{4 m E_{T}}\left(E_{T}+m\right)}$ and
$c_{s}=\frac{E_{T}+m}{\sqrt{4 m E_{T}}}$.

The odd one-photon terms can be eliminated by applying another transformation $e^{S_{s}^{\prime}}$. The transformed Hamiltonian can be obtained by using the expanded formula
\begin{equation}\label{FWTEboson}\begin{aligned}
H_{s}^{\prime \prime} =&U^{\dagger}\left(H_{s}^{\prime}-i \eta \partial_{t}\right) U \\
=&H_{s}^{\prime}+\sum_{n=1} \frac{1}{n !}\{S^{(n)},\left(H_{s}^{\prime}-i \eta \partial_{t}\right)\}+\ldots \\
=&H_{s}^{\prime}+\{S,\left(H_{s}^{\prime}-i  \eta \partial_{t}\right)\}
+\frac{1}{2}\{S,\{S,\left(H_{s}^{\prime}-i  \eta \partial_{t}\right)\}\}+\ldots.
\end{aligned}\end{equation}
Its one order term satisfies $\{S_{s}^{\prime}, E_{T}- i \eta \partial_{t}\}+O^{\prime}=0$, where $S_{s}^{\prime}$ can be derived by the recurrence relation,
\begin{equation}
S_{s}^{\prime}=-\frac{1}{2 m}\sum_{n=0}\left\{\left(\frac{i \eta \partial_{t}-(\sqrt{\pi^{2}+m^{2}}-m)}{2 m}\right)^{(n)}, O^{\prime}\right\}.
\end{equation}
The transformed Hamiltonian is
\begin{equation}
H^{\prime \prime}_{s}=E_{T}+H_{1 \gamma}^{\prime}+H_{2 \gamma}^{\prime}+\ldots,
\end{equation}
where $H_{2 \gamma}^{\prime}=\frac{1}{2}\left\{S_{s}^{\prime}, O^{\prime}\right\}$, and higher-order terms are neglected.

The nonrelativistic Hamiltonian up to $m\alpha^8$ order can be obtained by expanding $H^{\prime \prime}_{s}$ on momentum. It is
\begin{equation}\begin{aligned}
H^{\prime \prime}_{s}=& E_{T}+\eta A_{0}+\frac{i}{32} \eta\left[\pi^{2},\left\{\pi^{i}, E^{i}\right\}\right]\\
&-\frac{i}{32} \eta\left[\pi^{4},\left\{\pi^{i}, E^{i}\right\}\right]
-\frac{1}{32}\left\{\pi^{i}, E^{i}\right\}\left\{\pi^{i}, E^{i}\right\}
+o\left(m \alpha^{8}\right).
\end{aligned}\end{equation}
The first three terms are Hamiltonian up to $m\alpha^6 $. This result is coincided with the part without the form factor and multi-pole terms in Ref.~\cite{PhysRevA.82.052520}. The second line is $m\alpha^8$ order Hamiltonian.

\subsection{Spin-1 case} \label{spin1}
The vector field equation in the electromagnetic with minimal coupling is
$\pi_{\nu}\left(\pi^{\nu} u^{\mu}-\pi^{\mu} u^{\nu}\right)-m^{2} u^{\mu}=0$, where $u^\mu$ represents the vector field. The positive energy part and the negative part of the particle are
$\theta^{i}=\frac{1}{2}\left(u^{i}+\frac{1}{m}\left(\pi^{0} u^{i}-\pi^{i} u^{0}\right)\right)$
and
$\chi^{i}=\frac{1}{2}\left(u^{i}-\frac{1}{m}\left(\pi^{0} u^{i}-\pi^{i} u^{0}\right)\right)$ \cite{PhysRevA.82.052520}.
In the rest reference frame, positive state satisfies  $\chi^{i}=0$, and  negative state satisfies $\theta^{i}=0$. The index $i$ is the polarization direction of the vector field.

The vector field is
$\Psi^{i}=\left(\begin{array}{l}\theta^{i} \\ \chi^{i}\end{array}\right)$. And its equation of motion is $i \eta \partial_{t} \Psi^i=H_{V}^{ij} \Psi^{j}$, where the Hamiltonian is
\begin{equation}
H_{V}^{ij}=m\delta^{ij}+\frac{1}{2 m}\left(\vec{\pi}^{2}\delta^{ij}-\vec{\Sigma}^{ij} \cdot \vec{B}\right)
-\frac{1}{2 m} \tau\left(\vec{\pi}^{2}\delta^{ij}+\vec{\Sigma}^{ij} \cdot \vec{B}-2((\vec{\Sigma} \cdot \vec{\pi})^{2})^{ij}\right)
+\eta A^{0}\delta^{ij},
\end{equation}
$\vec B$ is the magnetic field, and $(\Sigma^i)^{jk}=-i\varepsilon^{ijk}$ is the spin operator of the vector field. Because spin is internal degrees, the equation and Hamiltonian can be expressed as
$i \eta \partial_{t} \Psi=H_{V} \Psi$
and
\begin{equation}\begin{aligned}
H_{V}=&m+\frac{1}{2 m}\left(\vec{\pi}^{2}-\vec{\Sigma} \cdot \vec{B}\right)
-\frac{1}{2 m} \tau\left(\vec{\pi}^{2}+\vec{\Sigma} \cdot \vec{B}-2(\vec{\Sigma} \cdot \vec{\pi})^{2}\right)+\eta A^{0}\\
=&\left(m+\frac{1}{2 m} p^{2}\right)+\frac{1}{2 m} \tau\left(-p^{2}+2(\vec{\Sigma} \cdot \vec{p})^{2}\right)+V_{E}+V_{O}.
\end{aligned}\end{equation}
In the second line, the first two terms are Hamiltonian of free particle, and the last two terms are even  and odd interactions
\begin{equation}\begin{array}{l}
V_{E}=\eta A^{0}+\frac{1}{2 m}\left(-\left\{p^{i}, A^{i}\right\}+\vec{A}^{2}-\vec{\Sigma} \cdot \vec{B}\right), \\
V_{O}=\frac{1}{2 m} \tau\left(\left\{p^{i}, A^{i}\right\}-\vec{A}^{2}-\vec{\Sigma} \cdot \vec{B}+2(\vec{\Sigma} \cdot \vec{A})^{2}-2\{\vec{\Sigma} \cdot \vec{p},\vec{\Sigma} \cdot \vec{A}\}\right).
\end{array}\end{equation}

For spin-1 particle, the relativistic Hamiltonian will be expanded with electromagnetic potential $A^\mu$.  The transformations of wave function and Hamiltonian are $\Psi=U\Psi^{\prime}$ and $H^{\prime}_{FW}=U^{\dagger}\left(H-i \eta \partial_{t}\right) U$.
The $U=e^{S}$ satisfy $U^{\dagger} \eta U=\eta$, and $S$ is Hermitian operator.  After performing the transformation 
$e^{S}=c_{V}+s_{V}$, where   
$s_{V}=\frac{\tau\left[p^{2}-2(\Sigma \cdot p)^{2}\right]}{2 \sqrt{m E_{T}}\left(m+E_{T}\right)}$,
$c_{V}=\frac{\left(E_{T}+m\right)}{\sqrt{4 m E_{T}}}$ and $E_{T}=\sqrt{p^2+m^2}$,
the transformed Hamitonian is 
\begin{equation}\begin{aligned}
&H_{V}^{\prime}=e^{S_{V}}\left(H_{V}-i \eta \partial_{t}\right) e^{S_{V}}
=E_{T}+H_{1\gamma}+O^{\prime},
\end{aligned}\end{equation}
and one-photon terms are 
\begin{equation}\begin{aligned}
&H_{1\gamma}=c_{V} V_{E} c_{V}+s_{V} V_{E} s_{V}+c_{V} V_{O} s_{V}+s_{V} V_{O} c_{V},\\
&O^{\prime}=c_{V} V_{O} c_{V}+s_{V} V_{O} s_{V}+c_{V} V_{E} s_{V}+s_{V} V_{E} c_{V}.
\end{aligned}\end{equation}
The odd one-photon terms $O^{\prime}$ can be cancelled by using Eq.(\ref{FWTEboson}). The  $S^{\prime}_{V}$ is
\begin{equation}
S_{V}^{\prime}=-\frac{1}{2 m}\sum_{n=0}\left\{\left(\frac{i \eta \partial_{t}-(\sqrt{p^{2}+m^{2}}-m)}{2 m}\right)^{(n)},O^{\prime}\right\}.
\end{equation}
The second transformed Hamiltonian is
\begin{equation}
H^{\prime \prime}_{V}=E_{T}+H_{1 \gamma}^{\prime}+H_{2 \gamma}^{\prime}+\ldots
\end{equation}
where $H_{2 \gamma}^{\prime}=\frac{1}{2}\left\{S^{\prime}_{V}, O^{\prime}\right\}$ is the two-photons term. And higher-order terms are neglected.

The FW Hamiltonian up to $m\alpha^8$ order can be obtained by expanding $H^{\prime \prime}_{V}$ with momentum. It is
\begin{equation}
\begin{aligned}
H_{V}^{\prime \prime} &=E_{T}+\eta A_{0}-\frac{1}{2}\left(\left\{p^{i}, A^{i}\right\}+\widetilde{B}\right) 
+\frac{1}{16} \eta\left[\left(p^{2}-2 \widetilde{p}^{2}\right),\left[\left(p^{2}-2 \widetilde{p}^{2}\right), A_{0}\right]\right] \\
&+\frac{1}{8}\left\{\left(\left\{p^{i}, A^{i}\right\}-\widetilde{B}-2\{\widetilde{p}, \widetilde{A}\}\right),\left(p^{2}-2 \widetilde{p}^{2}\right)\right\}
+\frac{1}{2} A^{2}\\
&-\eta \frac{1}{16}\left[\left(p^{2}-2 \widetilde{p}^{2}\right) p^{2},\left[\left(p^{2}-2 \widetilde{p}^{2}\right), A_{0}\right]\right]
-\frac{1}{64}\left\{p^{4},\left(\left\{p^{i}, A^{i}\right\}+\widetilde{B}\right)\right\} \\
&-\frac{1}{32}\left(p^{2}-2 \widetilde{p}^{2}\right)\left(\left\{p^{i}, A^{i}\right\}+\widetilde{B}\right)\left(p^{2}-2 \widetilde{p}^{2}\right) \\
&-\frac{1}{16}\left\{\left(\left\{p^{i}, A^{i}\right\}-\widetilde{B}-2\{\widetilde{p}, \widetilde{A}\}\right),\left(p^{2}-2 \widetilde{p}^{2}\right) p^{2}\right\}\\
&-\frac{1}{8}\left(\left\{p^{i}, A^{i}\right\}-\widetilde{B}-2\{\widetilde{p}, \widetilde{A}\}\right)^{2}+\frac{1}{32}\left[p^{2}-2 \widetilde{p}^{2}, A^{0}\right]\left[p^{2}-2 \widetilde{p}^{2}, A^{0}\right]\\
&+\frac{1}{8}\left\{\left(A^{2}+2 \widetilde{A}^{2}\right),\left(p^{2}-2 \widetilde{p}^{2}\right)\right\} \\
&-\frac{1}{16} \eta\left[\left(\left\{p^{i}, A^{i}\right\}-\widetilde{B}-2\{\widetilde{p}, \widetilde{A}\}\right),\left[p^{2}-2 \widetilde{p}^{2}, A^{0}\right]\right]+o\left(m \alpha^{8}\right),
\end{aligned}
\end{equation}
where the tilde is defined as $\widetilde{A}= A \cdot \Sigma$. The first two line is Hamiltonian up to $m\alpha^6$ order, and the rest are $m\alpha^8$ order terms.

\section{Elimination of singularities}\label{ElimS}

The nonrelativistic Hamiltonians of Coulomb Bound state are expanded on $p^2_{a}$, $V_{C}$ or equivalent term $\alpha^2\sim\langle p^2_{a} \rangle$, 
\begin{equation} 
\begin{aligned}
H=H_{0}+H_{(4)}+H_{(6)}+H_{(8)}+o(m\alpha^8).
\end{aligned}
\end{equation}
$H_{0}$ is the Sc$\ddot{o}$dinger Hamiltonian, which is  $m\alpha^2$ order. $H_{(4)}$, $H_{(6)}$ and $H_{(8)}$ are $m\alpha^4$,$m\alpha^6$ and $m\alpha^8$ order. The relativistic corrections are calculated by using the perturbation theory. At the $m\alpha^4$ order, it is $E_{(4)}=\langle H_{(4)}\rangle$. The expected value $H_{(4)}$ is finite for Coulomb Bound state, and $E_{(4)}$ can be calculated directly. 
   
The  $m\alpha^6$ and $m\alpha^8$ order energy corrections are 
\begin{equation} 
\begin{aligned}
E_{(6)}=\langle H_{(6)}\rangle+\langle H_{(4)}\frac{Q}{E_{0}-H_{0}}H_{(4)}\rangle,
\end{aligned}
\end{equation}
and
\begin{equation} 
\begin{aligned}
E_{(8)}=&\langle H_{(8)}\rangle+\langle H_{(6)}\frac{Q}{E_{0}-H_{0}}H_{(4)}\rangle
+\langle H_{(4)}\frac{Q}{E_{0}-H_{0}}H_{(6)}\rangle
\\&
+\langle H_{(4)}\frac{Q}{E_{0}-H_{0}}(H_{(4)}-E_{{4}})\frac{Q}{E_{0}-H_{0}}H_{(4)}\rangle,
\end{aligned}
\end{equation}
where $Q=I-P,P=|\varphi\rangle\langle\varphi|$ and $\varphi$ is the perturbed state. The singularities will appear in these terms, and their expected values are infinite. For example, the total contribution of $H_{(6)}$ contains $\langle \dfrac{\delta^3(r)}{r}\rangle$ and $\langle \dfrac{1}{r^{4}}\rangle$ terms. They are divergent for s states of hydrogen. The second perturbation term $\langle H_{(4)}\frac{Q}{E_{0}-H_{0}}H_{(4)}\rangle$ is also divergent. These divergent parts should be cancelled exactly. Because the nonrelativistic Hamiltonian is unitary equivalent to the relativistic Hamiltonian, which has finite eigenvalues. It is essential to separate and cancel the singular terms from original nonrelativistic Hamiltonian, and then deriving the calculable formula.

Some conventions are adopted to simplify the derivation. Two operators are defined as $A=E_0-V_C$ and $B=H_0-E_0$. $E_0,V_C$ are energy of perturbed state and the Coulomb interaction. Then the operators $H_{(4)},H_{(6)}$ and $H_{(8)}$ are expanded by the following formula 
\begin{equation}\label{HBexpanding}
\begin{aligned}
H_{(n)}=H^{00}_{(n)}+H^{01}_{(n)}B+BH^{10}_{(n)}+BH^{11}_{(n)}B,
\end{aligned}
\end{equation} 
where $n=4,6,8$.

Because $\varphi$ is the perturbed state, $B|\varphi\rangle=(H_0-E_0)|\varphi\rangle=0$. The leading order energy correction is 
\begin{equation}  \label{E4}
\begin{aligned}
E_{(4)}=\langle H_{(4)}\rangle=\langle H_{(4)}^{00}\rangle,
\end{aligned}
\end{equation}
and the $m\alpha^6$ order energy correction is equal to 
\begin{equation}  \label{E6}
\begin{aligned}
E_{(6)}=\langle H_{(6)}'\rangle-\langle H^{00}_{(4)}\frac{Q}{B}H^{00}_{(4)}\rangle,
\end{aligned}
\end{equation}
where
\begin{equation}\label{H6p}
\begin{aligned}
H_{(6)}'\equiv H^{00}_{(6)}-H^{01}_{(4)}Q(H^{00}_{(4)}-E_{(4)})-(H^{00}_{(4)}-E_{(4)})QH^{10}_{(4)}-H^{01}_{(4)}BH^{10}_{(4)}.
\end{aligned}
\end{equation}
The last three term is separated from the $\langle H_{(4)}\frac{Q}{E_{0}-H_{0}}H_{(4)}\rangle$. If all the singular terms in $H_{(6)}'$ can be are cancelled. Then other terms in $E_{(6)}$ are finite.

At $m\alpha^8$ order, the relativistic correction is equal to 
\begin{equation}\label{E8}
\begin{aligned}
E_{(8)}=&\left\langle H_{(8)}'\right\rangle 
-\langle H_{(6)}' \frac{Q}{B} H_{(4)}^{00}\rangle 
-\langle H_{(4)}^{00} \frac{Q}{B} H_{(6)}'\rangle
+\langle H_{(4)}^{00} \frac{Q}{B}\left(H_{(4)}^{00}-E_{(4)}\right) \frac{Q}{B} H_{(4)}^{00}\rangle,
\end{aligned}
\end{equation}
where
\begin{equation}\label{H8p}
\begin{aligned}
H_{(8)}'\equiv& H_{(8)}^{00}-H_{(6)}^{01} Q H_{(4)}^{00}- H_{(6)}^{00} Q H_{(4)}^{10}- H_{(6)}^{01} B H_{(4)}^{10}
- H_{(4)}^{01} Q H_{(6)}^{00}- H_{(4)}^{00} Q H_{(6)}^{10}- H_{(4)}^{01} B H_{(6)}^{10}
\\&
+\left(H_{(4)}^{00}+H_{(4)}^{01} B\right) Q H_{(4)}^{11} Q\left(H_{(4)}^{00}+B H_{(4)}^{10}\right)+ H_{(4)}^{01} Q\left(B H_{(4)}^{10}+H_{(4)}^{01} B\right) Q H_{(4)}^{10} 
\\&
+ H_{(4)}^{00} Q H_{(4)}^{10} Q H_{(4)}^{10}+ H_{(4)}^{01} Q\left(H_{(4)}^{00}-E_{(4)}\right) Q H_{(4)}^{10}+ H_{(4)}^{01} Q H_{(4)}^{01} Q H_{(4)}^{00}.
\end{aligned}
\end{equation}
It maybe contains singularities in $H_{(8)}'$. We can expand $H_{(6)}'$ to cancel the singularities by applying a similar strategy. In order to illustrate this method is effective, we will study the Hamiltonians of scalar and fermion.

\subsection{Scalar particle in Coulomb field}

The $m\alpha^8$ order Hamiltonian of scalar particle in Coulomb field obtained in Sec.II is 
\begin{equation}\begin{aligned}
H_{s}=& \frac{p^{2}}{2}+ V_{C}-\frac{p^{4}}{8}
+\frac{p^{6}}{16}+\frac{1}{32}\left[p^{2},\left[p^{2},V_{C}\right]\right]\\
&-\frac{5p^{8}}{128}-\frac{1}{32} \left[p^{4},\left[p^{2},V_{C}\right]\right]
+\frac{1}{32}\left[p^{2},V_{C}\right]\left[p^{2},V_{C}\right]
+o\left(m \alpha^{8}\right),
\end{aligned}\end{equation}
where $V_C$ is the Coulomb potential. The relativistic corrections to Hamiltonian are 
\begin{equation} \begin{aligned}
H_{(4)}=-\frac{p^{4}}{8}
= -\frac{1}{2}A^{2} -\frac{1}{2}B^{2} -\frac{1}{2}\{A,B\},
\end{aligned}\end{equation}
\begin{equation}\begin{aligned}
H_{(6)}=&\frac{p^{6}}{16}+\frac{1}{32}\left[p^{2},\left[p^{2},V_{C}\right]\right]
\\=&
\frac{1}{2}A^{3} +\frac{1}{4}ABA+\frac{5}{8}\{A^{2},B\}+\frac{3}{8}\{B^{2},A\}
+\frac{3}{4}BAB+\frac{1}{2}B^{3},
\end{aligned}\end{equation}
and
\begin{equation}\begin{aligned}
 H_{(8)}=& -\frac{5p^{8}}{128}
-\frac{1}{32} \left[p^{4},\left[p^{2},V_{C}\right]\right]
+\frac{1}{32}\left[p^{2},V_{C}\right]\left[p^{2},V_{C}\right]
\\ \simeq & H_{(8)}^{00}=
-\frac{5}{8}A^{4} -\frac{3}{8}(ABA^{2}+A^{2}BA)-\frac{1}{4}AB^{2}A.
\end{aligned}\end{equation}
where the $p^{2}=2(A+B)$ is used. The $H_{(n)}\simeq  H_{(n)}^{00}$ means   $\langle H_{(n)} \rangle = \langle H_{(n)}^{00}\rangle$. The difference between $H_{(8)}$ and $H_{(8)}^{00}$ doesn't contribute to the $m\alpha^8$ order energy. 

The relativistic corrections to the energy can be derived by using the Eq.(\ref{E4})(\ref{E6})(\ref{H6p})(\ref{E8})(\ref{H8p})
\begin{equation}\label{E4s}
\begin{aligned}
E_{(4)}=-\dfrac{1}{2}\langle A^{2}\rangle,
\end{aligned}
\end{equation}

\begin{equation}\label{E6s} 
\begin{aligned}
E_{(6)}=&\langle H_{(6)}'\rangle-\langle H^{00}_{(4)}\frac{Q}{B}H^{00}_{(4)}\rangle
\\=&
\dfrac{1}{2}\langle A^{2}\rangle \langle A\rangle
-\dfrac{1}{4}\langle A^{2}\dfrac{Q}{B} A^{2}\rangle
,
\end{aligned}
\end{equation}
and
\begin{equation}\label{E8s}
\begin{aligned}
E_{(8)}=&\left\langle H_{(8)}'\right\rangle 
-\langle H_{(6)}' \frac{Q}{B} H_{(4)}^{00}\rangle 
-\langle H_{(4)}^{00} \frac{Q}{B} H_{(6)}'\rangle
+\langle H_{(4)}^{00} \frac{Q}{B}\left(H_{(4)}^{00}-E_{(4)}\right) \frac{Q}{B} H_{(4)}^{00}\rangle
\\=&
-\dfrac{1}{2}\langle A^{2}\rangle \langle A\rangle^2
-\dfrac{1}{8}\langle A^{2}\rangle^2
+\dfrac{1}{4}\langle A^{2}\dfrac{Q}{B} A^{2}\rangle \langle A\rangle
+\dfrac{1}{4}\langle A\dfrac{Q}{B} A^{2}\rangle \langle A^{2}\rangle
+\dfrac{1}{4}\langle A^{2}\dfrac{Q}{B} A\rangle \langle A^{2}\rangle
\\&
-\dfrac{1}{8}\langle A^{2}\dfrac{Q}{B} (A^{2}+2E_{(4)})\dfrac{Q}{B} A^{2}\rangle 
\\=&
-E_{(4)}^2+E_{(2)}E_{(6)}+\dfrac{1}{2}\langle A\dfrac{Q}{B} A^{2}\rangle \langle A^{2}\rangle
-\dfrac{1}{8}\langle A^{2}\dfrac{Q}{B} (A^{2}+2E_{(4)})\dfrac{Q}{B} A^{2}\rangle ,
\end{aligned}
\end{equation}
where the regularized Hamiltonians are 
$H^{00}_{(4)}=-\dfrac{1}{2}A^{2}$, 
$H_{(6)}'=\dfrac{1}{4}(A^{2}PA+APA^{2})+\dfrac{1}{2}E_{(4)}(AP+PA)-E_{(4)}A$, 
and 
$H_{(8)}'=-\dfrac{1}{2}APA^{2}PA-\dfrac{1}{8}A^{2}PA^{2}$. All the terms in the $E_{(4)}, E_{(6)}$, and $E_{(8)}$ are finite. We will give some numerical results in the next section.

\subsection{Electron in Coulomb field}

The Hamiltonian of electron in Coulomb field is
\begin{equation} 
\begin{aligned}
H=& \frac{p^{2}}{2}+V_{C}
-\frac{p^{4}}{8}
-\frac{1}{8}[\widetilde{p},[\widetilde{p},V_{C}]]
\\&
+\frac{p^{6}}{16}
+\frac{3}{64}\{\widetilde{p}^2,[\widetilde{p},[\widetilde{p},V_{C}]]\}
+\frac{5}{128}[\widetilde{p}^2,[\widetilde{p}^2,V_{C}]
-\frac{1}{16}\{[\widetilde{p},V_{C}],[\widetilde{p},V_{C}]\}
\\&
-\frac{5p^{8}}{128}
-\frac{5}{128}\{\widetilde{p}^4,[\widetilde{p},[\widetilde{p},V_{C}]]\}
+\frac{9}{1024}[\widetilde{p}^2,[\widetilde{p}^2,
[\widetilde{p},[\widetilde{p},V_{C}]]]]
-\frac{19}{512}[\widetilde{p}^4,[\widetilde{p}^2,V_{C}]
\\&
+\frac{1}{64}\{[\widetilde{p},V_{C}],
\{\widetilde{p},\{\widetilde{p},[\widetilde{p},V_{C}]\}\}\}
+\frac{3}{64}\{[\widetilde{p},V_{C}],\{p^2,[\widetilde{p},V_{C}]\}\}
+o(m\alpha^8),
\end{aligned}
\end{equation}
$\widetilde{p}=\vec{p}\cdot\vec{\sigma}$, and $V_C$ is the Coulomb potential. The last two terms in the first line is $m\alpha^4$ order. It is easy to check that they are spin-orbit coupling terms and Darwin terms. The second line is $m\alpha^6$ order terms and third and fourth lines are  $m\alpha^8$ order terms. It is the Hamiltonian of hydrogen-like atoms or Hydrogen molecular ion. They are different in potential $V_C$. The specific form of $V_C$ are not applied in this section. 

By using $A=E_{0}-V_C$ and $B=H_0-E_0$, the relativistic Hamiltonian $H_{(n)} (n=4,6,8)$ are transformed to the $H^{00}_{(n)}+H^{01}_{(n)}B+BH^{10}_{(n)}+BH^{11}_{(n)}B
$ by applying Eq.(\ref{HBexpanding}). The $m\alpha^4$ order Hamiltonian is 
\begin{equation} \label{H4f}
\begin{aligned}
H_{(4)}=-\frac{p^{4}}{8}
-\frac{1}{8}[\widetilde{p},[\widetilde{p},V_{C}]]
= -\frac{1}{4}\widetilde{p}V_{C}\widetilde{p} 
-\frac{1}{4}\{A,B\} -\frac{1}{2}B^{2},
\end{aligned}
\end{equation}
and
\begin{equation} 
\begin{aligned}
H_{(4)}^{00}=&-\frac{1}{4}\widetilde{p}V_{C}\widetilde{p},
\\ 
H_{(4)}^{01}=&H_{(4)}^{10}= -\frac{1}{4}A,
\\
H_{(4)}^{11}=&-\frac{1}{2}.
\end{aligned}
\end{equation}
The $m\alpha^6$ order Hamiltonian is 
\begin{equation} 
\begin{aligned}
H_{(6)}=&\frac{p^{6}}{16}
+\frac{3}{64}\{\widetilde{p}^2,[\widetilde{p},[\widetilde{p},V_{C}]]\}
+\frac{5}{128}[\widetilde{p}^2,[\widetilde{p}^2,V_{C}]
-\frac{1}{16}\{[\widetilde{p},V_{C}],[\widetilde{p},V_{C}]\}
\\
=&
-\frac{1}{4}A^{3} +\frac{1}{16}ABA-\frac{1}{32}\{A^{2},B\}
+\frac{5}{32}\{B^{2},A\}
+\frac{7}{16}BAB+\frac{1}{2}B^{3}
\\&
+\frac{3}{16}(\widetilde{p}A\widetilde{p}A+A\widetilde{p}A\widetilde{p}
+\widetilde{p}A\widetilde{p}B+B\widetilde{p}A\widetilde{p}),
\end{aligned}
\end{equation}
and
\begin{equation} 
\begin{aligned}
H_{(6)}^{00}=&-\frac{1}{4}A^{3} +\frac{1}{16}ABA
+\frac{3}{16}(\widetilde{p}A\widetilde{p}A+A\widetilde{p}A\widetilde{p}),
\\ 
H_{(6)}^{01}=&-\frac{1}{32}A^{2}
+\frac{5}{32}AB+\frac{3}{16}\widetilde{p}A\widetilde{p},
\\
H_{(4)}^{10}=&-\frac{1}{32}A^{2}
+\frac{5}{32}BA+\frac{3}{16}\widetilde{p}A\widetilde{p} ,
\\
H_{(6)}^{11}=&\frac{7}{16}A+\frac{1}{2}B.
\end{aligned}
\end{equation}
The $m\alpha^8$ order Hamiltonian
\begin{equation}
\begin{aligned}
H_{(8)}=&-\frac{5p^{8}}{128}
-\frac{5}{128}\{\widetilde{p}^4,[\widetilde{p},[\widetilde{p},V_{C}]]\}
+\frac{9}{1024}[\widetilde{p}^2,[\widetilde{p}^2,
[\widetilde{p},[\widetilde{p},V_{C}]]]]
-\frac{19}{512}[\widetilde{p}^4,[\widetilde{p}^2,V_{C}]
\\&
+\frac{1}{64}\{[\widetilde{p},V_{C}],
\{\widetilde{p},\{\widetilde{p},[\widetilde{p},V_{C}]\}\}\}
+\frac{3}{64}\{[\widetilde{p},V_{C}],\{p^2,[\widetilde{p},V_{C}]\}\}
\\
\simeq & H_{(8)}^{00},
\end{aligned}
\end{equation}
and
\begin{equation}\label{H800f}
\begin{aligned}
H_{(8)}^{00}=&\frac{A^{4}}{4}+\frac{37}{128}\left(A B A^{2}+A^{2} B A\right)+\frac{21}{64} A B^{2} A
\\&
-\frac{3}{16} \widetilde{p} A^{3} \widetilde{p}+\frac{3}{32}\left(\widetilde{p} A^{2} \widetilde{p} A+A \widetilde{p} A^{2} \widetilde{p}\right) 
-\frac{19}{128}\left(\widetilde{p} A \widetilde{p} A^{2}+A^{2} \widetilde{p} A \widetilde{p}\right)
\\&
-\frac{1}{32}\left(\widetilde{p} B A^{2} \widetilde{p}+\widetilde{p} A^{2} B \widetilde{p}\right)-\frac{27}{128}(\widetilde{p} A \widetilde{p} B A+A B \widetilde{p} A \widetilde{p}) 
\\&
+\frac{1}{32}(\widetilde{p} B A \widetilde{p} A+A \widetilde{p} A B \bar{p})+\frac{5}{32}(\tilde{p} A B \widetilde{p} A+A \widetilde{p} B A \widetilde{p})
\\&
-\frac{9}{64} A \widetilde{p} A \widetilde{p} A-\frac{3}{16} A \widetilde{p} B \widetilde{p} A-\frac{1}{8} \widetilde{p} A B A \widetilde{p},
\end{aligned}
\end{equation}
The terms $H_{(8)}^{01},H_{(8)}^{10}$ and $H_{(8)}^{11}$ are neglected. Because $\langle H^{01}_{(n)}B+BH^{10}_{(n)}+BH^{11}_{(n)}B \rangle
=0$, their contributions to $m\alpha^8$ order energy correction is zero.

The relativistic corrections to the energy can be derived by using the Eq.(\ref{E4})(\ref{E6})(\ref{H6p})(\ref{E8})(\ref{H8p}). The $m\alpha^4$ order energy correction is  
\begin{equation} \label{E4f}
\begin{aligned}
E_{(4)}=\langle H_{(4)}\rangle=\langle H_{(4)}^{00}\rangle=-\dfrac{1}{4}\langle\widetilde{p} A\widetilde{p}\rangle,
\end{aligned}
\end{equation}
At the $m\alpha^6$ order, the energy corrections can be obtained by using the Eq.(\ref{E6}) (\ref{H6p})  
\begin{equation} 
\begin{aligned}
E_{(6)}=\langle H_{(6)}'\rangle-\langle H^{00}_{(4)}\frac{Q}{B}H^{00}_{(4)}\rangle,
\end{aligned}
\end{equation}
The regularized $H_{(6)}'$ is
\begin{equation}\label{H6pF} 
\begin{aligned}
H_{(6)}'= & H^{00}_{(6)}-H^{01}_{(4)}Q(H^{00}_{(4)}-E_{(4)})-(H^{00}_{(4)}-E_{(4)})QH^{10}_{(4)}-H^{01}_{(4)}BH^{10}_{(4)}.
\\
= &\frac{1}{8}\widetilde{p}A^{2}\widetilde{p}
+\dfrac{1}{16}(\widetilde{p}A\widetilde{p}PA+AP\widetilde{p}A\widetilde{p})
-\frac{1}{4}(QA+AQ)E_{(4)}
+\frac{1}{8}\{A^{2},B\}
\end{aligned}
\end{equation}
where the equation $\widetilde{p}A\widetilde{p}A+A\widetilde{p}A\widetilde{p}=\widetilde{p}A^2\widetilde{p}+2A^3+\{B,A^2\}$ (Eq.(\ref{A1})) is applied, and all singular terms, such as the $A^3, ABA, \widetilde{p}A\widetilde{p}A$ and $A\widetilde{p}A\widetilde{p}$ in $H^{00}_{(6)}$ are cancelled.  The regularized $m\alpha^6$ order energy is 
\begin{equation} \label{E6f}
\begin{aligned}
E_{(6)}=&\dfrac{1}{8}\langle\widetilde{p} A^{2}\widetilde{p}\rangle
+\dfrac{1}{8}\langle\widetilde{p} A\widetilde{p}\rangle\langle A \rangle
-\dfrac{1}{16}
\langle \widetilde{p} A\widetilde{p}\dfrac{Q}{B} \widetilde{p} A\widetilde{p}\rangle
\end{aligned}
\end{equation}
All the terms in the equation is finite.

The $m\alpha^8$ order the energy correction is derived by using Eq.(\ref{E8}),(\ref{H8p}). However, these terms are still divergent for electron. Define $H_{(6)}'^{00}= H_{(6)}'-\dfrac{1}{8}\{A^{2}, B\}$. The  energy correction $E_{(8)}$ in Eq.(\ref{E8}) is equal to
\begin{equation}
\begin{aligned}
E_{(8)}
=&\left\langle H_{(8)}''\right\rangle 
-\langle H_{(6)}'^{00} \frac{Q}{B} H_{(4)}^{00}\rangle 
-\langle H_{(4)}^{00} \frac{Q}{B} H_{(6)}'^{00}\rangle
+\langle H_{(4)}^{00} \frac{Q}{B}\left(H_{(4)}^{00}-E_{(4)}\right) \frac{Q}{B} H_{(4)}^{00}\rangle,
\end{aligned}
\end{equation}
and the regularized $H_{(8)}''$ is
\begin{equation}
\begin{aligned}
H_{(8)}''\equiv& H_{(8)}'-\dfrac{1}{8}A^{2} Q H_{(4)}^{00}- \dfrac{1}{8} H_{(4)}^{00} Q A^{2}.
\end{aligned}
\end{equation}
Almost all singular terms except $\widetilde{p}A^3\widetilde{p}$ in  $H_{(8)}''$ can be cancelled by using the equations  $\widetilde{p}A^2\widetilde{p}A+A\widetilde{p}A^2\widetilde{p}=\dfrac{4}{3}\widetilde{p}A^3\widetilde{p}+\dfrac{4}{3} A^4+\dfrac{2}{3}\{B,A^3\}$ and $\widetilde{p}A\widetilde{p}A^2+A^2\widetilde{p}A\widetilde{p}=\dfrac{2}{3}\widetilde{p}A^3\widetilde{p}+\dfrac{8}{3} A^4+\dfrac{4}{3}\{B,A^3\}$ (Eq.(\ref{A1})). 

Then, the $m\alpha^8$ order the energy correction is
\begin{equation} \label{E8f}
\begin{aligned}
E_{(8)}=&-\frac{1}{64}\left\langle\widetilde{p} A \widetilde{p} \frac{Q}{B}\left( \widetilde{p} A \widetilde{p}+4E_{(4)}\right) \frac{Q}{B} \widetilde{p} A \widetilde{p}\right\rangle
+\frac{1}{32}\langle
\widetilde{p} A^{2} \widetilde{p} \frac{Q}{B} \widetilde{p} A \widetilde{p}+
\widetilde{p} A \widetilde{p} \frac{Q}{B} \widetilde{p} A^{2} \widetilde{p}\rangle 
-\frac{1}{16} \langle\widetilde{p} A^{3} \widetilde{p} \rangle
\\&
-\frac{1}{8}E_{(4)}
\langle A \frac{Q}{B} \widetilde{p} A \widetilde{p}+\widetilde{p} A \widetilde{p} \frac{Q}{B} A \rangle
-E_{(4)}^2+\frac{1}{2}E_{(2)}E_{(6)}.
\end{aligned}
\end{equation}
All the terms in the second line are finite. In the first line, each term is divergent for s state. A further regularized procedure should be applied. And the first three terms in the first line should be calculated together. By applying $\langle\widetilde{p} A^{3} \widetilde{p} \rangle=\langle\widetilde{p} A^{2}QA \widetilde{p} \rangle=\langle\widetilde{p} AQAQA \widetilde{p} \rangle$, the second and third terms in the first line can be absorbed into first terms. The $m\alpha^8$ energy correction is 
\begin{equation}  \label{E8f1}
\begin{aligned}
E_{(8)}=&-\frac{1}{64}\left\langle\widetilde{p} A 
(\widetilde{p} \frac{Q}{B}\widetilde{p}-2Q) A 
(\widetilde{p} \frac{Q}{B} \widetilde{p}-2Q)
A\widetilde{p}\right\rangle
+\frac{1}{64}\langle\widetilde{p} A \widetilde{p} \rangle \left\langle\widetilde{p} A \widetilde{p} \frac{Q}{B^2}\widetilde{p} A \widetilde{p}\right\rangle
\\&
-\frac{1}{8}E_{(4)}
\langle A \frac{Q}{B} \widetilde{p} A \widetilde{p}+\widetilde{p} A \widetilde{p} \frac{Q}{B} A \rangle
-E_{(4)}^2+\frac{1}{2}E_{(4)}E_{(6)}.
\end{aligned}
\end{equation}
All the divergent parts are combined in the first term. The results is finite. We will give some numerical results of hydrogen in next section.

\subsection{Multi-electron systems in Coulomb field}
The Hamiltonian of multi-electron systems in Coulomb field is 
\begin{equation} 
\begin{aligned}
H=& \sum_{a}\left(\frac{p^{2}_{a}}{2}\right)+V_{C}
+\sum_{a}\left(-\frac{p^{4}_{a}}{8}
-\frac{1}{8}[\widetilde{p_{a}},[\widetilde{p_{a}},V_{C}]]\right)
\\&
+\sum_{a}\left(\frac{p^{6}_{a}}{16}
+\frac{3}{32}[\widetilde{p_{a}}^3,[\widetilde{p_{a}},V_{C}]]
-\frac{1}{128}[\widetilde{p_{a}}^2,[\widetilde{p_{a}}^2,V_{C}]
-\frac{1}{16}\{[\widetilde{p_{a}},V_{C}],[\widetilde{p_{a}},V_{C}]\}
\right)
\\&
+\sum_{a<b}\left(
\frac{1}{64}[\widetilde{p_{a}},[\widetilde{p_{a}},[\widetilde{p_{b}},[\widetilde{p_{b}},V_{C}]]]]
\right)
+o(m\alpha^6),
\end{aligned}
\end{equation}
where the Coulomb interactions 
\begin{equation} 
\begin{aligned}
V_{C}=-\sum_{i,a}\frac{Z_{i}}{r_{ia}}+\sum_{a<b}\frac{1}{r_{ab}},
\end{aligned}
\end{equation}
contains both the electron-electron (subscript i,a) and electron-nucleus interaction (subscript a,b).

Relativistic correction to Hamiltonian $H_{(4)}$ and $H_{(6)}$ are transformed by the same method. The  $m\alpha^4$ order corrections to Hamiltonian is  
\begin{equation} 
\begin{aligned}
H_{(4)}=&\sum_{a}\left(-\frac{p^{4}_{a}}{8}
-\frac{1}{8}[\widetilde{p_{a}},[\widetilde{p_{a}},V_{C}]]\right)
\\=&
 -\frac{1}{4}\sum_{a}\widetilde{p}_{a}V_{C}\widetilde{p}_{a}
+\dfrac{1}{4}\sum_{a<b} p^{2}_{a}p^{2}_{b}
-\frac{1}{4}\{A,B\} -\frac{1}{2}B^{2},
\end{aligned}
\end{equation}
and
\begin{equation} 
\begin{aligned}
H_{(4)}^{00}=&
-\frac{1}{4}\sum_{a}\widetilde{p}_{a}V_{C}\widetilde{p}_{a}
+\dfrac{1}{4}\sum_{a<b} p^{2}_{a}p^{2}_{b},
\\ 
H_{(4)}^{01}=&H_{(4)}^{10}= -\frac{1}{4}A,
\\
H_{(4)}^{11}=&-\frac{1}{2}.
\end{aligned}
\end{equation}
The $m\alpha^6$ order correction to Hamiltonian
\begin{equation} 
\begin{aligned}
H_{(6)}=&\sum_{a}\left(\frac{p^{6}_{a}}{16}
+\frac{3}{32}[\widetilde{p_{a}}^3,[\widetilde{p_{a}},V_{C}]]
-\frac{1}{128}[\widetilde{p_{a}}^2,[\widetilde{p_{a}}^2,V_{C}]
-\frac{1}{16}\{[\widetilde{p_{a}},V_{C}],[\widetilde{p_{a}},V_{C}]\}
\right)
\\&
+\sum_{a<b}\left(
\frac{1}{64}[\widetilde{p_{a}},[\widetilde{p_{a}},[\widetilde{p_{b}},[\widetilde{p_{b}},V_{C}]]]]
\right)
\\=&
\sum_{a}\left(\frac{p^{6}_{a}}{16}
+\frac{3}{64}\{\widetilde{p}^2_{a},[\widetilde{p}_{a},[\widetilde{p}_{a},V_{C}]]\}
+\frac{1}{32}(q_{a}V_{C})^{2}\right)
\\&
+\sum_{a<b}\left(
\frac{1}{64}[\widetilde{p_{a}},[\widetilde{p_{a}},[\widetilde{p_{b}},[\widetilde{p_{b}},V_{C}]]]]
+\frac{5}{64}[p_{a}^{2},[p_{b}^{2},V_{C}]]
\right)
+\frac{5}{32}[H_{0},[H_{0},V_{C}]]]
\end{aligned}
\end{equation}
and 
\begin{equation}
\begin{aligned}
H_{(6)}^{00}=&-\frac{1}{4} A^{3}+\frac{1}{16} A B A+\frac{3}{16} \sum_{a<b<c} p_{a}^{2} p_{b}^{2} p_{c}^{2}+\frac{3}{16}\left\{A, \sum_{b} \tilde{p}_{b} A \widetilde{p}_{b}-\sum_{a<b} p_{a}^{2} p_{b}^{2}\right\} \\
&-\sum_{a \neq b}\left(-\frac{3}{64}\left\{\hat{p}_{a}^{2},\left\{\hat{p}_{b}^{2}, A\right\}\right\}+\frac{3}{128}\left[\hat{p}_{a}^{2},\left[\hat{p}_{b}^{2}, A\right]\right]+\frac{3}{32}\left\{\hat{p}_{a}^{2}, \widetilde{p}_{b} A \tilde{p}_{b}\right\}\right) \\
&-\sum_{a<b} \frac{1}{64}\left[\tilde{p}_{a},\left[\widetilde{p}_{a},\left[\tilde{p}_{b},\left[\widetilde{p}_{b}, A\right]\right]\right]\right].
\end{aligned}
\end{equation}
The terms $H_{(6)}^{01},H_{(6)}^{10}$ and $H_{(6)}^{11}$ are neglected.

The $m\alpha^4$ order energy correction is 
\begin{equation}
\begin{aligned}
E_{(4)}=\langle H_{(4)} \rangle=\langle H_{(4)}^{00} \rangle=
-\frac{1}{4} \sum_{a}
\langle \widetilde{p}_{a}V_{C}\widetilde{p}_{a}\rangle
+\dfrac{1}{4}\sum_{a<b} \langle p^{2}_{a}p^{2}_{b}\rangle
\end{aligned}
\end{equation}
At $m\alpha^6$ order, the terms $\langle H_{(4)} \frac{Q}{E-H_{0}} H_{(4)}\rangle$ and $\langle  H_{(6)}\rangle$ are divergent. The divergent parts are
\begin{equation}
\begin{aligned}
\langle H_{(4)} \frac{Q}{E-H_{0}} H_{(4)}\rangle=&
\langle H_{(4)}^{00} \frac{Q}{E-H_{0}} H_{(4)}^{00}\rangle
+\frac{1}{16}\langle\{\left(- \sum_{a} \tilde{p}_{a} A \widetilde{p}_{a}^{i}+ \sum_{a<b} p_{a}^{2} p_{b}^{2}\right), A\}\rangle \\
&-\frac{1}{16}\langle A B A\rangle
-\frac{1}{2}\langle H_{(4)}^{00}\rangle\langle A\rangle
\\
=&\frac{1}{16}\langle\{\left(- \sum_{a} \tilde{p}_{a} A \widetilde{p}_{a}+ \sum_{a<b} p_{a}^{2} p_{b}^{2}\right), A\}\rangle 
-\frac{1}{16}\langle A B A\rangle+finite,
\end{aligned}
\end{equation}
and 
\begin{equation}
\begin{aligned}
\langle H_{(6)}^{00}\rangle=&-\frac{1}{4} \langle A^{3} \rangle+\frac{1}{16} \langle A B A \rangle
+\frac{3}{16}\langle
\left\{A, \sum_{b} \tilde{p}_{b} A \widetilde{p}_{b}-\sum_{a<b} p_{a}^{2} p_{b}^{2}\right\}\rangle+finite.
\end{aligned}
\end{equation}
All the divergence can be cancelled by using the equations 
 $\sum_{a}(\widetilde{p}_{a}A\widetilde{p}_{a}A+A\widetilde{p}_{a}A\widetilde{p}_{a})=\sum_{b}\widetilde{p}_{a}A^2\widetilde{p}_{a}+2A^3+\{B,A^2\}$ (Eq.(\ref{A1})), and the total contribution of $m\alpha^6$ order energy is 
\begin{equation}
\begin{aligned}
E_{(6)}=
&\frac{1}{8} \sum_{a}\left\langle\widetilde{p}_{a} A^{2} \widetilde{p}_{a}\right\rangle
-\frac{1}{8} \sum_{a<b}\left\langle\left\{A, p_{a}^{2} p_{b}^{2}\right\}\right\rangle
+\frac{3}{16} \sum_{a<b<c}\left\langle p_{a}^{2} p_{b}^{2} p_{c}^{2}\right\rangle 
\\&
-\sum_{a<b}\left(
-\frac{3}{32}\left\langle\left\{p_{a}^{2},\left\{p_{b}^{2}, A\right\}\right\}\right\rangle+\frac{3}{64}\left\langle\left[p_{a}^{2},\left[p_{b}^{2}, A\right]\right]\right\rangle
+\frac{3}{16}\left\langle\left\{p_{a}^{2}, \widetilde{p}_{b} A \widetilde{p}_{b}\right\}\right\rangle\right) 
\\&
-\sum_{a<b} \frac{1}{64}\left\langle\left[\widetilde{p}_{a},\left[\tilde{p}_{a},\left[\tilde{p}_{b},\left[\tilde{p}_{b}, A\right]\right]\right]\right]\right\rangle+\langle H_{(4)}^{00} \frac{Q}{E-H_{0}} H_{(4)}^{00}\rangle-\frac{1}{2}\left\langle H_{(4)}^{00}\right\rangle\langle A\rangle.
\end{aligned}
\end{equation}
All the terms are finite. It can be proved by the following reason. The divergence always appears when one electron approaches the nucleus. It is obvious the first term is finite for s state of Slater-type functions. The summation of power of $p^2_{a}$ and $\frac{1}{r_{a}}$ in the formula except first term is less than three. All terms are finite when $r_{a}\rightarrow 0$. It applies to other electrons.

\section{The relativistic correction of scalar particle and electron in Coulomb field }\label{checking}

We will check the formula deriving in this work by comparing their expectation values with the analytic relativistic result of scalar particle and electron in Coulomb field. Because the leading order terms of nonrelativistic Hamiltonian of scalar particle and electron are the same. The eigenvalues, eigenstates and Green function of nonrelativistic hydrogen \cite{Swainson_1991a,Swainson_1991b,Swainson_1991c} are used to derive the relativistic correction. 

The eigenfunction of hydrogen atom is
\begin{equation}
\psi_{n l m}(r, \theta, \phi)=R_{n l}(r) Y_{l m}(\theta, \phi),
\end{equation}
where $R_{n l}$ is radial part and $ Y_{l m}$ is spherical harmonics function. The radial part of the nonrelativistic wave function can be written as 
\begin{equation}
R_{n l}(r)=N_{n l}(2 r / n)^{l} \mathrm{e}^{-r / n} L_{n-l-1}^{2 l+1}(2 r / n),
\end{equation}
$ L_n^{l}(z) $ is the generalized Laguerre function and $N_{n l}=\left(2 / n^{2}\right) \sqrt{(n-l-1) ! /(n+l) !}$ is the normalization factor. 

In coordinate representation, the operator $Q/B$ is reduced Green function. It can be expressed as
\begin{equation}
G^{\prime}\left(\boldsymbol{r}_{1}, \boldsymbol{r}_{2} ; E_{n}\right)
=\left.\langle\boldsymbol{r}_{1}\right|\dfrac{Q}{B}\left|\boldsymbol{r}_{2}\rangle\right.
=\sum_{k,l^{\prime} m^{\prime}} g_{l^{\prime},k}\left(r_{1}, r_{2} ; E_{n}\right) Y_{l^{\prime} m^{\prime}}\left(\theta_{1}, \phi_{1}\right) 
Y_{l^{\prime} m^{\prime}}^{*}\left(\theta_{2}, \phi_{2}\right).
\end{equation}
where the radial function can be expressed as 
\begin{equation}
\begin{aligned}
g_{l^{\prime},k\neq n-l^{\prime}-1}\left(r_{1}, r_{2}; E_{n}\right)=
\left(\frac{2}{n}\right)^{2 l^{\prime}+1} 
\frac{k !2\left(r_{1} r_{2}\right)^{l^{\prime}} 
\mathrm{e}^{-\left(r_{1}+r_{2}\right) / n}}
{\left(2 l^{\prime}+1+k\right) !\left(l^{\prime}+1+k-n\right)} L_{k}^{2 l^{\prime}+1}\left(\frac{2 r_{1}}{n}\right) L_{k}^{2 l^{\prime}+1}\left(\frac{2 r_{2}}{n}\right),
\end{aligned}
\end{equation}
and 
\begin{equation}
\begin{aligned}
g_{l^{\prime},n-l^{\prime}-1}\left(r_{1}, r_{2}; E_{n}\right) &=
2\left(\frac{2}{n}\right)^{2 l^{\prime}+1}\left(r_{1} r_{2}\right)^{l^{\prime}} 
\mathrm{e}^{-\left(r_{1}+r_{2}\right) / n} 
\frac{\left(n-l^{\prime}-1\right) !}{2 n\left(n+l^{\prime}\right) !}
\\& 
\times
\left\{
\left[\left(n-l^{\prime}\right) L_{n-l^{\prime}}^{2 l^{\prime}+1}\left(\frac{2 r_{2}}{n}\right)-\left(n+l^{\prime}\right) L_{n-l^{\prime}-2}^{2 l^{\prime}+1}\left(\frac{2 r_{2}}{n}\right)\right] 
L_{n-l^{\prime}-1}^{2 l^{\prime}+1}\left(\frac{2 r_{1}}{n}\right) \right.
\\&
+\left[\left(n-l^{\prime}\right) L_{n-l^{\prime}}^{2 l^{\prime}+1}\left(\frac{2 r_{1}}{n}\right)-\left(n+l^{\prime}\right) L_{n-l^{\prime}-2}^{2 l^{\prime}+1}\left(\frac{2 r_{1}}{n}\right)\right]
L_{n-l^{\prime}-1}^{2 l^{\prime}+1}\left(\frac{2 r_{2}}{n}\right)
\\&
+\left.L_{n-l^{\prime}-1}^{2 l^{\prime}+1}\left(\frac{2 r_{1}}{n}\right) L_{n-l^{\prime}-1}^{2 l^{\prime}+1}\left(\frac{2 r_{2}}{n}\right)
\right\}.
\end{aligned}
\end{equation}

Although reduced Green function is a summation of infinite terms. The term 
\begin{equation}\label{B1}
\begin{aligned}
\langle H^{00}_{(4)}\dfrac{Q}{B}H^{00}_{(4)}\rangle=\sum_{k=0}^{\infty}D_{k} ,
\end{aligned}
\end{equation}
in second perturbation of relativistic energy ($m\alpha^6$ order) can be calculated by the following two method. (1)Analytical method: The first ten terms are calculated accurately. Then the general term formulas are conjectured. The general term formulas are tested by comparing with exact results of first hundred terms, and the summation can be obtained accurately. (2)Approximate method: The first $\Omega$ terms are calculated accurately. For $\Omega\gg 1$, $D_{k}\simeq \dfrac{c_{0}}{k^{a}}+\dfrac{c_{1}}{k^{a+1}}+...$, the coefficient $c_0,c_1...$ and $a$ are extracted from the first $\Omega$ terms. Then the approximated result is    
\begin{equation}
\begin{aligned}
\langle H^{00}_{(4)}\dfrac{Q}{B}H^{00}_{(4)}\rangle=\sum_{k=0}^{\infty}D_{k}\simeq \sum_{k=0}^{\Omega}D_{k}+\sum_{i=0}^{q}\sum_{k=\Omega+1}^{\infty}\dfrac{c_i}{k^{a+i}}.
\end{aligned}
\end{equation}
The accuracy could be improved by increasing the $q$. In the following calculation, $q=2$ is chosen. The third perturbation term like $\langle H^{00}_{(4)}\dfrac{Q}{B}H^{00}_{(4)}\dfrac{Q}{B}H^{00}_{(4)}\rangle$ can be calculated by the similar methods. Almost all the corrections can be obtained by two method, except $m\alpha^8$ order relativistic correction of the s state of election in Coulomb field. It will be calculated by applying the first method.

The relativistic energy of scalar particle in Coulomb field  $V_C=-\frac{1}{r}$ is 
\begin{equation}
\begin{aligned}
E_{nl}=&\left(1+\dfrac{\alpha^2}{(n-l-1/2+\sqrt{(l+1/2)^2-\alpha^2})^{2}}\right)^{-1/2}
\\=&
 1-\dfrac{\alpha^2}{2n^2}+\left(3-\dfrac{4n}{(l+1/2)}\right)\dfrac{\alpha^6}{8n^4}+....,
\end{aligned}
\end{equation}
where the $n,l$ are the principal and orbital angular momentum quantum number. The ground state energy $E_{10}= 1-\dfrac{1}{2}\alpha^2-\dfrac{5}{8}\alpha^4-\dfrac{21}{16}\alpha^6-\dfrac{429}{128}\alpha^8+o(\alpha^{8})$. The first two terms in the Taylor expansion are mass and nonrelativistic energy. Others are high-order relativistic corrections. 

Here, we will test the relativistic corrections Eq.(\ref{E4s})(\ref{E6s})(\ref{E8s}). The leading order of relativistic corrections Eq.(\ref{E4s}) can be derived analytically,
\begin{equation}
\begin{aligned}
E_{4}=-\dfrac{1}{2}\langle A^{2}\rangle =-\dfrac{1}{2}\left( E_{(2)}^2
+2E_{(2)}\langle \dfrac{1}{r}\rangle
+\langle \dfrac{1}{r^2}\rangle\right)
=\dfrac{3}{8n^4}-\dfrac{1}{2n^3(l+1/2)}.
\end{aligned}
\end{equation}
It is coincided with the third terms of the Taylor expansion of the relativistic energy $E_{nl}$. The relativistic corrections (\ref{E6s})(\ref{E8s}) are calculated by the methods mentioned before. The results are given in the Table I-III. $E_6$ is obtained by applying the approximate method with $\Omega=100, q=2$. Comparing with the exact vale, they have at least eight significant figures. The $E_6$ and $E_8$ of $1s$ state obtained by applying the analytical method are coincided with the exact value.

\begin{table}[!h] 
\caption{The approximate $E_{(6)}$ of $1s$ state of scalar in Coulomb field.}
\begin{tabular}{ll} 
\hline  $\Omega$   & ~~~~~~~~$E_{(6)}$ \\ \hline\hline
10 & -1.3124906365\\[3pt]  
20 & -1.3124990733\\[3pt]
30 & -1.3124997838\\ [3pt] 
40 & -1.3124999255\\ [3pt] 
50 & -1.3124999678\\[3pt]  
60 & -1.3124999839\\ [3pt] 
70 & -1.3124999911\\ [3pt]   
80 & -1.3124999947\\ [3pt]   
90 & -1.3124999966\\ [3pt]   
100& -1.3124999978\\ [3pt]   
$\infty$& -1.3124999993(15)\\ [3pt]
Exact   &-1.3125\\ [3pt]    
\hline   
\end{tabular}
\end{table}

\begin{table}[!h] 
\caption{The  $E_{(6)}$ of  scalar in Coulomb field. The second column is the calculated by applying the approximate method. There are extrapolating results from $\Omega=100$. }
\begin{tabular}{lll} 
\hline  States   & Approximate & Exact \\ \hline\hline
$1s$ & -1.312499999(2) &-1.3125\\[3pt] 
$2s$ & -0.17675781250(2) &-0.1767578125\\[3pt]  
$2p$ &-0.0043041085(8)& -0.0043041087...\\[3pt]
$3s$ &-0.049811385(1)& -0.049811385...\\ [3pt] 
$3p$ &-0.0018004112(8)& -0.0018004115...\\ [3pt] 
$3d$ &-0.000231138548(7)& -0.000231138545...\\[3pt]    
\hline   
\end{tabular}
\end{table}

\begin{table}[!h] 
\caption{The relativistic energy of $1s$ state of scaler in Coulomb field.
The first four line are calculated by applying the analytical method mentioned below Eq.(\ref{B1}). Then the relativistic energy Eq.(\ref{E4s})(\ref{E6s})(\ref{E8s}) are derived. They are coincided with the Taylor expansion of the relativistic energy.}
\begin{tabular}{cc} 
\hline    & Exact   \\ \hline\hline
  &  \\  
$ \langle A^{2}\rangle$ & $\frac{5}{4}$ \\[3pt]  
$ \langle A^{2}\dfrac{Q}{B} A^{2}\rangle$ & $\frac{13}{2} $\\[3pt]
$ \langle A\dfrac{Q}{B} A^{2}\rangle $ & $\frac{3}{2}$   \\ [3pt] 
$ \langle A^{2}\dfrac{Q}{B} (A^{2}+2E_{(4)})\dfrac{Q}{B} A^{2}\rangle$ &\scriptsize$38$\scriptsize\\ [3pt] 
$ E_4$ & $-\frac{5}{8}$  \\[3pt]  
$ E_6$ & $-\frac{21}{16}$  \\ [3pt] 
$ E_8$ & $-\frac{429}{128}$  \\ [3pt]     
\hline   
\end{tabular}
\end{table}

The relativistic energy of hydrogen is
\begin{equation}
\begin{aligned}
E_{n\kappa}=&\left(1+\dfrac{\alpha^2}{(n-\left|\kappa\right| +\sqrt{\kappa^2-\alpha^2})^{2}}\right)^{-1/2}
\\=&
1-\dfrac{\alpha^2}{2n^2}+\left(3-\dfrac{4n}{\left|\kappa\right|}\right)\dfrac{\alpha^6}{8n^4}+...,
\end{aligned}
\end{equation}
where $\kappa=\pm(j+1/2)$. If $\kappa>0$, $l=\kappa-1$ and if $\kappa<0$, $l=-\kappa$. $n,j,l$ are principal, total angular momentum and orbital angular momentum quantum number. In the second line, the first two terms in the Taylor expansion are mass and nonrelativistic energy. The third term is leading order relativistic corrections. The higher-order relativistic corrections can be obtained from Taylor expansion. The energy of ground state is $E_{11}=\sqrt{1-\alpha^2}= 1-\dfrac{1}{2}\alpha^2-\dfrac{1}{8}\alpha^4-\dfrac{1}{16}\alpha^6-\dfrac{5}{128}\alpha^8+o(\alpha^{8})$.
 
In Table.IV-V, approximate results of relativistic correction Eq.(\ref{E6f})(\ref{E8f}) are listed. At $m\alpha^6$ order, the approximate results have at least eight significant figures. At $m\alpha^8$ order, there is artificial divergence appear in relativistic correction of the $s$ state. The approximate method is invalid. However, all the expectation values of the operators of relativistic corrections are finite for $l>0$ state. Approximate results have at least five significant figures for $E_8$.

The relativistic correction Eq.(\ref{E4f})(\ref{E6f})(\ref{E8f}) of ground state of hydrogen are calculated by applying analytical method. The result are listed in Table.VI. Some expectation values are divergent. The divergent parts are proportional to a Harmonic series. However, the divergent part in $\langle\widetilde{p} A^{3} \widetilde{p}\rangle$ can't be separate from the finite part directly. The key point is applying the equation $2\langle\widetilde{p} A^{3} \widetilde{p}\rangle=\langle\widetilde{p} A^{2}\dfrac{Q}{B}BA \widetilde{p}\rangle+\langle\widetilde{p} AB\dfrac{Q}{B}A^{2} \widetilde{p}\rangle$. Then a divergent Harmonic series appears. All the divergent Harmonic series cancelled in the total contributions $E_8$. The relativistic corrections up $m\alpha^8$ order are coincided with the exact value.

\begin{table}[htb] 
\caption{The $E_{(6)}$ of hydrogen. The second column is extrapolating result from $\Omega=100$.}
\begin{tabular}{lll} 
\hline  States~~ & Approximate~~ & Exact \\ \hline\hline
$1s$ & -0.0624999998(4) &-0.0625\\[3pt] 
$2s$ & -0.020507812501(4) &-0.0205078125\\[3pt]  
$2p$ &-0.0009765623(5)&-0.0009765625...\\[3pt]
$3s$ &-0.0066015090(3)& -0.0066015089...\\ [3pt] 
$3p$ &-0.0006215704(4)& -0.0006215706...\\ [3pt] 
$3d$ &-0.000085733884(5)& -0.000085733882...\\[3pt]    
\hline   
\end{tabular}
\end{table}

\begin{table}[htb] 
\caption{Approximate $E_{(8)}$ of hydrogen. The second column is extrapolating result from $\Omega=40$.}
\begin{tabular}{ccc} 
\hline   States~~ & Approximate~~ & Exact
 \\   \hline\hline  
$2p$ &  -0.00015256(6)  & -0.00015258... \\ [3pt]
$ 3p$ & -0.00010029(6)  & -0.00010032...\\  [3pt]
$ 3d$ & -0.0000059538(4)&  -0.0000059537... \\  [3pt]    
\hline   
\end{tabular}
\end{table}

\begin{table}[!h] 
\caption{Some exact expectation value of $1s$ state of hydrogen. They are calculated by applying the analytical method mentioned below Eq.(\ref{B1}). The $h_s=\sum \frac{1}{i}$ is Harmonic series. The fine-structure constant is chosen as unit one here.}
\begin{tabular}{cc} 
\hline       & Exact value   \\ \hline\hline
$ \langle A\rangle$ & $\frac{1}{2}$ \\[3pt]  
$\langle\widetilde{p}A\widetilde{p}\rangle$& $\frac{1}{2}$ \\[3pt]  
$ \dfrac{1}{8}\langle\widetilde{p} A^{2}\widetilde{p}\rangle$ & $\frac{5}{32} $\\[3pt]
$\dfrac{1}{16}\langle\widetilde{p} A\widetilde{p}\dfrac{Q}{B} \widetilde{p} A\widetilde{p} \rangle$ 
& $-\frac{1}{4}$   \\ [3pt]   
$ -\frac{1}{16}\langle\widetilde{p} A^{3} \widetilde{p}\rangle$ 
& $\frac{67}{128}-\frac{1}{4}h_s$   \\ [3pt] 
$\frac{1}{32}\langle\widetilde{p} A^{2} \widetilde{p} \frac{Q}{B} \widetilde{p} A \widetilde{p}+
\widetilde{p} A \widetilde{p} \frac{Q}{B} \widetilde{p} A^{2} \widetilde{p}\rangle$  
& $-\frac{13}{32}+\frac{1}{2}h_s$  \\[3pt]
$-\frac{1}{64}\langle\widetilde{p} A \widetilde{p} \frac{Q}{B}\left( \widetilde{p} A \widetilde{p}+4E_{(4)}\right) \frac{Q}{B} \widetilde{p} A \widetilde{p}\rangle$ 
& $-\frac{3}{16}-\frac{1}{4}h_s$  \\ [3pt] 
$\langle A \frac{Q}{B} \widetilde{p} A \widetilde{p}+\widetilde{p} A \widetilde{p} \frac{Q}{B} A\rangle$  & $-\frac{1}{4}$  \\ [3pt]
 $E_{(4)}$ in Eq.(\ref{E4f}) &  $-\frac{1}{8}$ \\[3pt] 
 $E_{(6)}$ in Eq.(\ref{E6f}) &  $-\frac{1}{16}$ \\[3pt] 
 $E_{(8)}$ in Eq.(\ref{E8f}) &  $-\frac{5}{128}$ \\[3pt]     
\hline   
\end{tabular}
\end{table}

\section{Conclusion} \label{conclusion}
 
In this work, the nonrelativisitc Hamiltonians of spin-0,~1/2,~1 particles in the electromagnetic field are derived by applying DKH-FW method. The result of spin-1/2 is coincided with the previous result obtained by using scattering matching approach \cite{PhysRevA.100.012513}. Comparing with the scattering matching of NRQED, the even one-photon potential of FW Hamiltonian can be obtained by expanding the photon-fermion scattering amplitude. The even two-photon potential of FW Hamiltonian originated in the process of quantum electrodynamic: the positive-energy state can transit to negative-energy intermediate state and transits back to positive-energy state though emit/absorb two photons. It indicates that the two approaches up to $m\alpha^8$ order are equivalent at the tree-level, which can satisfy nowadays accuracy requirement. Although the result of spin-0 and spin-1 obtained by scattering matching approach is absent. It is easy to find that $H_{2 \gamma}^{\prime}$ is the product of the odd one-photon terms of DKH Hamiltonian and the Green function of negative-energy state. Their two-photon terms may be interpreted as the contribution of the negative energy state. Their equivalence should be proved in the further works.

Parts of high-order corrections to the energy of Coulombic systems can be derived by using these FW Hamiltonians. One is the self-energy correction of low-energy virtual photons. Another is the photon-exchange interactions between electrons or electron and nucleus. The Coulomb-photon-exchange interactions are studied in the work. And the nonrelativistic Coulomb Hamiltonians are obtained. Then, the equivalent formulas of energy corrections are derived. The singularities of Hamiltonian are cancelled. The numerical results of scalar and electron up to $m\alpha^8$ are coincided with the relativistic formula. The regularized Coulomb Hamiltonian can be using to calculate relativistic corrections of the Hydrogen molecular ion. Further more, it indicates the singularities in the tree, single loop and multi-loop Feynman diagram contribution or gauge invariant Feynman diagram may be canceled parts by parts. It is essential to simplify obtaining or verifying the higher-order corrections.

\appendix
\section{Some proof of equations}

By using $\sum_{a}\widetilde{p}_{a}^2=2(A+B)$ and $ [A,[\widetilde{p}_{a},A]]=0$, $\sum_{a}
(\widetilde{p}_{a}A^{m}\widetilde{p}_{a}A^n
+A^n\widetilde{p}_{a}A^{m}\widetilde{p}_{a})$ can be transformed to 
\begin{equation}
\begin{aligned}
\sum_{a}
(\widetilde{p}_{a}A^{m}\widetilde{p}_{a}A^n
+A^n\widetilde{p}_{a}A^{m}\widetilde{p}_{a})
=&\sum_{a}
(\widetilde{p}_{a},[A^{m},\widetilde{p}_{a}]A^n
+A^n[\widetilde{p}_{a},A^{m}]\widetilde{p}_{a}
+\{\widetilde{p}_{a}^2, A^{m+n}\})
\\
=&\sum_{a}
([\widetilde{p}_{a},[A^{m},\widetilde{p}_{a}]A^n])
+4A^{m+n}+2\{B, A^{m+n}\}
\\
=&-\sum_{a}
(m[\widetilde{p}_{a},[\widetilde{p}_{a},A]A^{m+n-1}])
+4A^{m+n}+2\{B, A^{m+n}\},
\end{aligned}
\end{equation}
and 
\begin{equation}
\begin{aligned}
\sum_{a}
(\widetilde{p}_{a}A^{m}\widetilde{p}_{a}A^n
+A^n\widetilde{p}_{a}A^{m}\widetilde{p}_{a})
=&\sum_{a}
(\widetilde{p}_{a}A^{m}[\widetilde{p}_{a},A^n]
+[A^n,\widetilde{p}_{a}]A^{m}\widetilde{p}_{a}
+2\widetilde{p}_{a} A^{m+n}\widetilde{p}_{a})
\\
=&\sum_{a}
([\widetilde{p}_{a}A^{m}[\widetilde{p}_{a},A^n]]
+2\widetilde{p}_{a} A^{m+n}\widetilde{p}_{a})
\\
=&\sum_{a}
(n[\widetilde{p}_{a},[\widetilde{p}_{a},A]A^{m+n-1}]
+2\widetilde{p}_{a} A^{m+n}\widetilde{p}_{a}).
\end{aligned}
\end{equation}
Eliminating the terms $[\widetilde{p}_{a},[\widetilde{p}_{a},A]A^{m+n-1}]$ on the right of these two equations. It is equal to
\begin{equation} \label{A1}
\begin{aligned}
\sum_{a}(\widetilde{p}_{a}A^{m}\widetilde{p}_{a}A^n
+A^n\widetilde{p}_{a}A^{m}\widetilde{p}_{a})
=\dfrac{2m}{m+n}\sum_{a}\widetilde{p}_{a}A^{m+n}\widetilde{p}_{a}
+\dfrac{4n}{m+n} A^{m+n+1}
+\dfrac{2n}{m+n}\{B,A^{m+n}\}.
\end{aligned}
\end{equation}
The equations used in this work is 
\begin{equation}
\begin{aligned}
\sum_{a}(\widetilde{p}_{a}A\widetilde{p}_{a}A
+A\widetilde{p}_{a}A\widetilde{p}_{a})
=\sum_{a}\widetilde{p}_{a}A^{2}\widetilde{p}_{a}
+2 A^{3}+\{B,A^{2}\},
\end{aligned}
\end{equation}
\begin{equation}
\begin{aligned}
\widetilde{p}A^2\widetilde{p}A+A\widetilde{p}A^2\widetilde{p}=\dfrac{4}{3}\widetilde{p}A^3\widetilde{p}+\dfrac{4}{3} A^4+\dfrac{2}{3}\{B,A^3\}
\end{aligned},
\end{equation}
and
\begin{equation}
\begin{aligned}
\widetilde{p}A\widetilde{p}A^2+A^2\widetilde{p}A\widetilde{p}=\dfrac{2}{3}\widetilde{p}A^3\widetilde{p}+\dfrac{8}{3} A^4+\dfrac{4}{3}\{B,A^3\}.
\end{aligned}
\end{equation}

\textbf{ACKNOWLEDGMENTS}

This work was supported by the National Natural Science Foundation of China (Nos.12074295 and 12104420). X.-S. Mei was also supported by the Strategic Priority Research Program of the Chinese Academy of Sciences (No. XDB21020200).

\bibliography{1.bib}

\end{document}